\RequirePackage{fix-cm}
\documentclass[smallextended]{svjour3}       % onecolumn (second format)
\smartqed  % flush right qed marks, e.g. at end of proof
\usepackage{longtable,tabularx}
\usepackage{hyperref}
\usepackage{graphicx}
\usepackage{latexsym}
\usepackage{comment}
\usepackage{natbib}
\usepackage{tikz}
\usepackage{amsmath}
\usepackage{amssymb}
\usetikzlibrary{decorations.pathreplacing}
\usepackage{caption}
\usepackage{subcaption}
\usepackage[margin=1.2in]{geometry}
\usepackage{arydshln}

\usepackage{soul} % for \hl{}

\begin{document}
\title{Design optimization of semiconductor manufacturing equipment using a novel multi-fidelity surrogate modeling approach}


\author{
    Bingran Wang$^1$$^*$\and
    Min Sung Kim$^2$$^*$\and 
    Taewoong Yoon$^2$\and
    Dasom Lee$^2$\and
    Byeong-Sang Kim$^2$\and
    Dougyong Sung$^2$\and
    John T. Hwang$^1$ 
}


\institute{Corresponding author: Bingran Wang \at \email{b6wang@ucsd.edu}\\
           % Min Sung Kim \at
           %    \email{kms0224.kim@samsung.com}\\            
           $^1$ Department of Mechanical and Aerospace Engineering, University of California San Diego, La Jolla, CA 92093, USA \\
           $^2$ Mechatronics Research, Samsung Electronics Co., Ltd., 1-1 Samsungjeonja-ro, Hwaseong-si, Gyeonggi-do 18448, Republic of Korea\\
           $^*$ These authors contributed equally to this work
}

\date{Received: date / Accepted: date}

\maketitle
\begin{abstract}
Careful design of semiconductor manufacturing equipment is crucial for ensuring the performance, yield, and reliability of semiconductor devices. Despite this, numerical optimization methods are seldom applied to optimize the design of such equipment due to %the complexity of the design space and 
the difficulty of obtaining accurate simulation models. In this paper, we address a practical and industrially relevant electrostatic chuck (ESC) design optimization problem by proposing a novel multi-fidelity surrogate modeling approach. 
The optimization aims to improve the temperature uniformity of the wafer during the etching process by adjusting seven parameters associated with the coolant path and embossing.
Our approach combines low-fidelity (LF) and high-fidelity (HF) simulation data to efficiently predict spatial-field quantities, even with a limited number of data points.
We use proper orthogonal decomposition (POD) to project the spatially interpolated HF and LF field data onto a shared latent space, followed by the construction of a multi-fidelity kriging model to predict the latent variables of the HF output field. In the ESC design problem, with hundreds or fewer data, our approach achieves a more than 10\% reduction in prediction error compared to using kriging models with only HF or LF data. Additionally, in the ESC optimization problem, our proposed method yields better solutions with improvements in all of the quantities of interest, while requiring 20\% less data generation cost compared to the HF surrogate modeling approach.
\end{abstract}

\section{Introduction}

The semiconductor industry is rapidly advancing, driven by intense global competition and a persistent push to innovate. 
Semiconductor devices are continuing to shrink in size while growing increasingly complex, presenting significant challenges in design and manufacturing (\cite{chien2011modelling, wong2020density}). 
It has become increasingly critical and challenging to mitigate this growing complexity while simultaneously improving yield, cost-effectiveness, performance, and scalability. 
Addressing these challenges necessitates the use of design optimization methods to navigate these intricate and unintuitive trade-offs. 

The design of semiconductor manufacturing equipment involves many coupled parameters, but accurately modeling the impact of these parameters on metrics such as yield and performance is challenging. 
Experimentally measuring these quantities is not only costly but also time-consuming, while predicting these quantities through high-fidelity simulations requires substantial computational resources and time. 
As a result, developing fast and reliable predictive models for design optimization is difficult, due to the limited availability of high-quality data. 
The compounding challenges of optimizing many coupled design parameters with sparse data for modeling the influence of these parameters are common to many design problems in the development of the semiconductor manufacturing equipment.

Here, we are specifically interested in a design problem involving the electrostatic chuck, where the goal is to maximize the  uniformity of the temperature field on the wafer surface during the etching process. 
The electrostatic chuck is used to hold the semiconductor wafer during multiple stages of semiconductor manufacturing, including not only etching but also deposition and lithography. 
It is critical to precisely hold the wafer position through an electrostatic force, but it is also critical to limit the wafer temperature while minimizing variability across the surface through a carefully designed cooling system (\cite{o1992impact,bubenzer1990plasma}). 
The wafer temperature field can be predicted with relatively high accuracy using a dynamic heat-transfer simulation of the wafer coupled to a fluid dynamics simulation of the coolant flow (considered here as the high-fidelity model), or with lower accuracy but significantly reduced computational cost using a steady-state heat-transfer simulation with simplified assumptions for the coolant flow (considered here as the low-fidelity model) (\cite{yoon2023heat}). 

This design problem presents several characteristics that make design optimization challenging. 
The first challenge is that both the high- and low-fidelity models have expensive evaluation times. 
The low-fidelity model has a roughly order-of-magnitude lower computation time, but its accuracy is not sufficient for effective design. 
The second challenge is that while the design space is moderate-dimensional, the state space (whose elements represent the discretized temperature field) is high-dimensional. 
Thus, any surrogate model used in place of the high- and low-fidelity models needs to model the entire temperature field, because the precise quantities of interest can differ across practical settings, in terms of both the spatial quantity (e.g., mean, maximum, standard deviation) and the localization (e.g., inner radial zone, outer radial zone). 
In practice, training a surrogate model to predict specific scalar quantities would require retraining frequently, whenever the application demands changes to the optimization objective or constraints, at significant manual effort and time.
Additionally, the problem formulated here is representative of many challenges faced in the semiconductor manufacturing process, where often hundreds of etching steps involved.
In each of these steps, similar design problems arise for the ESC, but under varying plasma conditions, chemical compositions, and radial frequency power. 
As a result, the ESC design must be optimized for each specific etching condition. This makes the data generation cost of the surrogate modeling approach we use particularly crucial, as it must be applied repeatedly across various etching processes.

Motivated by these challenges, this paper presents a novel methodology for electrostatic chuck design optimization based on multi-fidelity surrogate modeling and proper orthogonal decomposition (POD). 
This methodology, specifically focused on temperature uniformity maximization, has three steps. 
The first step applies POD to reduce the dimension of the state space by computing a singular value decomposition given snapshots of the state vectors (i.e., the discretized, steady-state temperature field) at different points in the parameter space. 
The second step constructs a surrogate model from the low-fidelity simulation data that predicts the discretized temperature field in the reduced state space. 
Since the outputs are a small set of coefficients with respect to the reduced basis, our methodology permits use of one of a wide range of possible surrogate modeling approaches, such as kriging. 
The third step constructs a second surrogate model from the high-fidelity simulation data that also predicts the discretized temperature field in the reduced state space, where this surrogate model is the discrepancy function in a multi-fidelity model.

We describe this novel methodology in the context of a specific, industrially relevant electrostatic chuck (ESC) design problem with seven design parameters, including emboss contact ratios and coolant channel dimensions. 
We report results found in applying the new methodology to this electrostatic chuck design problem, including the error due to the dimension reduction using POD, the accuracy of the multi-fidelity surrogate model compared to low- and high-fidelity surrogate models of roughly equivalent computational expense, and the validation of this methodology through evaluation of the computed optimized designs using the high-fidelity model, which is treated as the ground-truth model.

This paper is organized as follows: Section \ref{Sec: Background} provides an overview of surrogate-based design optimization, multi-fidelity surrogate modeling, and semiconductor manufacturing equipment optimization. Section \ref{Sec: Prob_Form} details the ESC design optimization problem and outlines the context and motivation for multi-fidelity surrogate modeling. 
Section \ref{Sec: Methodology} presents the proposed multi-fidelity surrogate modeling approach. Section \ref{Sec: Results} presents the surrogate modeling results and demonstrates the application of the proposed method to solve the ESC optimization problem. 
Finally, Section \ref{Sec: Conclusion} summarizes the findings and offers concluding remarks.

\section{Background}
\label{Sec: Background}
\subsection{Surrogate modeling in design optimization}

Surrogate models are widely used to approximate complex, expensive-to-evaluate functions in optimization problems. These methods are particularly useful in engineering design, where high-fidelity simulations, such as those based on computational fluid dynamics or finite element analysis, are costly in terms of both time and computational resources. Surrogate models, also known as \textit{metamodels}, provide a means of approximating the relationship between design variables and performance metrics with much lower computational expense.

Common surrogate models include radial basis functions (RBF) (\cite{buhmann2000radial}), kriging (\cite{krige1951statistical, matheron1963principles}), inverse distance weighting (IDW) (\cite{shepard1968two}), and neural networks (\cite{rumelhart1986learning}). Each of these models offers different advantages depending on the complexity and dimensionality of the problem. 
RBF constructs an approximation by employing a weighted sum of basis functions, each centered at known data points. The influence of a given data point decreases as the distance from the point of interest increases. RBF models are effective in capturing smooth, continuous functions, with the weights determined by solving a typically dense linear system. 
Kriging, also known as Gaussian process regression, generalizes the RBF approach by using correlation functions to model spatial relationships between data points, providing not only predictions but also estimates of uncertainty. Like RBF, kriging involves solving a linear system to compute the weights, but its probabilistic nature allows for improved accuracy. Kriging is especially powerful for low-dimensional problems with sparse data, where the number of training points is small to moderate (typically less than 1,000) and the dimensionality is manageable (up to $O(100)$ (\cite{hwang2018fast}). However, both kriging and RBF become computationally expensive with larger datasets because the cost of solving the linear system grows with the size of the training data.
IDW is a deterministic interpolation method.
Unlike kriging, which assumes a probabilistic model and incorporates spatial correlations, IDW is based purely on the distances between the known and unknown data points, assigning more weight to closer points and less weight to those farther away.
IDW can handle a very large amount of data (up to $10^5$) as it does not require any training. However, IDW is slow to predict especially when the training data is large, which makes it less favorable in design optimization problems.
Neural networks consist of layers of interconnected nodes (neurons), where each neuron applies a nonlinear transformation to the weighted sum of its inputs. 
This method can learn complex relationships between inputs and outputs and has been extremely powerful in handling high-dimensional problems in many applications such as computer vision (\cite{voulodimos2018deep}) and large language models (\cite{achiam2023gpt}).
However, their applicability in surrogate modeling for design optimization is often limited by the need for large datasets. 
% Neural networks typically require at least thousands of data points to outperform simpler models like RBF or kriging in terms of accuracy. 
% As a result, when the dataset is small, neural networks may underperform compared to other surrogate models, making them less favorable for most engineering problems, where data generation is often costly and time-consuming.

\subsection{Multi-fidelity surrogate modeling methods}
In many engineering problems, there are multiple computational models available to describe the system of interest, each differing in evaluation cost and fidelities. Typically, high-fidelity (HF) models, though computationally expensive, offer the accuracy necessary for critical applications, whereas low-fidelity (LF) models, though faster and less resource-intensive, provide less precise approximations. 
Multi-fidelity surrogate modeling methods aim to combine data from various fidelity levels, enabling the construction of surrogate models that leverage the strengths of both HF and LF models. 
This approach can significantly reduce the number of expensive HF evaluations required, achieving the efficient use of data from multiple sources.

One of the simplest approaches to multi-fidelity surrogate modeling involves using additive or multiplicative correction models~\cite{kennedy2001bayesian}. In an additive model, the LF model provides a baseline prediction, and a discrepancy function is learned from HF data to correct the LF model’s bias. In contrast, multiplicative correction models apply a scaling factor to the LF model, adjusting for systematic differences in magnitude between LF and HF models. While these methods are easy to implement and perform well when the LF model provides a reasonable approximation, they rely on the assumption that the relationship between LF and HF data is either additive or multiplicative. This assumption may not hold in complex systems with nonlinear interactions, limiting the effectiveness of these methods in such cases.

A more advanced and flexible approach is multi-fidelity kriging, also known as co-kriging (\cite{le2013multi}). Co-kriging is an extension of kriging that combines data from multiple fidelity levels into a single surrogate model. The LF data serves as a prior estimate, and the HF data is used to model the discrepancy between the two fidelity levels. This method is often recursive and can be particularly advantageous as it offers greater flexibility than simpler additive or multiplicative models. However, co-kriging can become computationally expensive, especially for high-dimensional problems or large datasets, as it requires solving multiple Gaussian process models. Its performance is also highly dependent on the correlation between LF and HF models; if the LF model is poorly correlated with the HF data, co-kriging may not significantly improve accuracy.

In recent years, machine learning techniques have been integrated into multi-fidelity modeling.
This approach relies on a large number of available data and trains the neural network models that can map both LF and HF data into a lower-dimensional latent space and reconstruct the HF predictions from this latent space. This allows for efficient surrogate modeling with high-dimensional field outputs. 
Deep neural networks and transfer learning approaches are among the popular techniques and have demonstrated superior performance in fluids (\cite{sun2020surrogate}) and aerodynamics (\cite{li2022deep,shen2024application}).
% In 
% One popular approach is transfer learning, where a surrogate model is first trained on abundant LF data and then fine-tuned using a smaller amount of HF data. This approach reduces the dependency on large amounts of HF data while maintaining accuracy. Another emerging method involves using autoencoders, which are neural networks that map both LF and HF data into a lower-dimensional latent space. The decoder reconstructs the HF predictions from this latent space, allowing for efficient surrogate modeling with high-dimensional data such as temperature fields or fluid flow simulations. 
These machine learning-based methods are highly flexible and can capture complex, non-linear interactions between LF and HF data, making them particularly useful in data-rich environments. However, they often require large training datasets and can be computationally expensive to train.

In fact, the performance of every multi-fidelity surrogate modeling method is dependent on the correlation between LF and HF models. If the LF model is poorly correlated with the HF data, the multi-fidelity surrogate models can easily be outperformed by the surrogate modeling methods using the HF data. More comprehensive reviews on multi-fidelity surrogate modeling methods can be found in (\cite{fernandez2016review,peherstorfer2018survey}).

\subsection{Semiconductor manufacturing equipment optimization}
% Optimizing semiconductor manufacturing equipment is essential for improving yield and performance as devices become increasingly complex.
The literature includes several studies in which optimization has been applied to semiconductor manufacturing equipment design.
Two specific application areas are deposition systems and electrostatic chucks (ESCs), both of which require precise control of parameters such as temperature and gas flow.
In deposition processes, the showerhead plays a crucial role in ensuring uniform gas distribution (\cite{liao2018modeling}). Recent research by (\cite{jin2024machine}) has focused on optimizing the showerhead design for better gas flow uniformity using machine learning-based optimization methods and computational fluid dynamics (CFD) simulations.  By adjusting geometric parameters like hole patterns and sizes, their machine learning-based framework achieved a 10\% improvement in gas flow uniformity but the machine learning framework required thousands of CFD evaluations.
Temperature control during processes like plasma etching and chemical vapor deposition (CVD) is another key challenge, and ESCs are critical in maintaining uniform wafer temperatures. Studies have focused on optimizing backside gas pressure and ceramic contact ratios to enhance heat transfer between the wafer and the ESC. In (\cite{yoon2023heat}), Yoon et al. demonstrated that optimizing gas pressure improved temperature uniformity up to an optimal threshold.
Youn and Hong (\cite{youn2024enhanced}) used CFD simulations to show that increasing the ceramic contact ratio improved heat transfer but introduced challenges such as localized non-uniformity.
A model developed by Klick and Bernt (\cite{klick2006microscopic}) explored the effects of different gas species on wafer cooling efficiency, offering insights into optimizing gas types like helium and neon for improved thermal management.

% \begin{itemize}
%     \item Surrogate-based modeling methods
%     \item Multi-fidelity methods
% \end{itemize}
% \textcolor{blue}{Descriptions on semiconductor process device. Etching process device (device figure). Parameterizations and why these parameters are important. Low-fidelity and high-fidelity simulations, what simulations models are used and what are not included in the low-fidelity simulation model (e.g. turbulence?). Why is it important to build the surrogate model to predict the entire temperature field and why uniformity is the only quantity considered in this paper. Optimization table.}
\section{Problem Formulation}
\label{Sec: Prob_Form}
\subsection{Electrostatic chuck design optimization problem}
Plasma is a physical state consisting of various particles such as electrons, positive/negative ions, and radicals. The compositions of these particles can be easily modified by changing the type of gas, and their behavior can be controlled through electromagnetic fields. Due to its high reactivity and chemical instability, plasma plays a crucial role in several semiconductor manufacturing processes, including modern etching, deposition, and cleaning.

\begin{figure}[h]
    \centering
    \includegraphics[width = 15 cm]{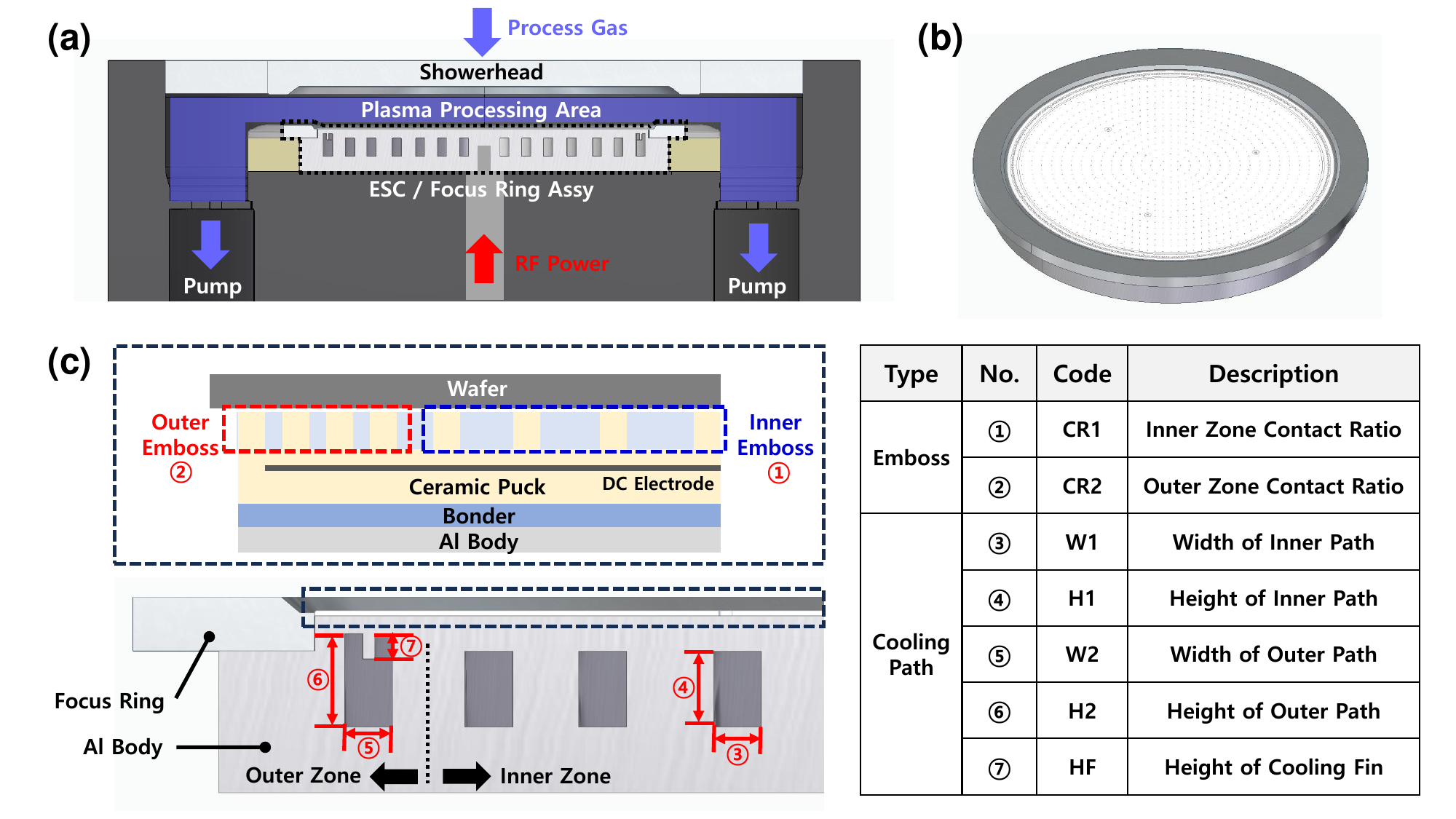}
    \caption{Electrostatic chuck (ESC) visualizations with indications of the seven design variables:  (a)  major components of capacitively coupled plasma etcher (b) 3D ESC assembly model (c) coolant path of ESC (d) surface emboss structure of ESC.
}
    \label{fig:ESC_devices}
\end{figure}

In the etching process, especially in high aspect ratio etching, maintaining precise etch profile control and uniformity is critical, particularly as devices are scaled down. The success or failure of semiconductor devices highly depends on this level of control. 
In response, advanced equipment technologies have been developed to regulate key process parameters, such as gas composition and flow rates, ion acceleration and tilting via radio frequency (RF) systems, and wafer temperature control to manage etch rate and uniformity~\cite{may2006fundamentals}. These systems are designed to control the plasma characteristics as well as the chemical and physical etching behaviors on the wafer surface.
Figure~\ref{fig:ESC_devices}~(a) presents a schematic diagram of a capacitive coupled plasma etcher, a system equipped with an upper electrode that includes a showerhead and a lower electrode that serves as an electrostatic chuck (ESC).
 The ESC, where the wafer is placed, plays a dual role. It not only enables physical/anisotropic etching through the application of bias RF power but also acts as a cooling system, managing the heat incident on the wafer and focus ring during the etching process. As device scaling progresses, etching requires higher bias power, increasing the demand for effective thermal management technology to dissipate the heat generated by the plasma. This has made cooling efficiency a critical area of development for ESC systems. Figures~\ref{fig:ESC_devices} (b), ~\ref{fig:ESC_devices} (c), and ~\ref{fig:ESC_devices} (d) show a 3D CAD model and cross-sectional views of the ESC and focus ring assembly, respectively.

The ESC assembly primarily consists of a ceramic puck, an aluminum body, and an insulating bonder that joins the two. Inside the aluminum body is a spiral-shaped coolant path, which is designed to dissipate heat transferred from the wafer and the focus ring. The efficiency of this cooling system is determined by several key parameters, including the path’s geometry, cross-sectional shape, and the flow rate of the coolant. A DC electrode embedded in the ceramic puck generates electrostatic forces to hold the wafer in place during the process. Additionally, backside helium (HE) gas is injected into the gap between the wafer and the ESC surface, formed by the emboss structures on the ceramic puck. The emboss contact ratio—the percentage of the wafer surface in direct contact with the ESC—along with backside HE gas pressure, plays a crucial role in controlling the wafer’s surface temperature, making it a key design variable.

In this paper, we focus on optimizing seven critical design parameters associated with the electrostatic chuck (ESC) process device, which plays a vital role in ensuring the quality of semiconductor manufacturing. 
The design variables include the parameters associated with the emboss contact ratio and the coolant path widths and heights.
The device, along with its corresponding design variables, can be visualized in Fig.~\ref{fig:ESC_devices}.
The objective of this optimization is to enhance the uniformity of the temperature field on the wafer surface during the etching process, as temperature uniformity is a key factor influencing the overall quality and yield of the semiconductor process. Specifically, we aim to minimize the three-sigma (3$\sigma$) value of the temperature field, which serves as a measure of its uniformity.

The optimization problem is formulated to minimize the 3$\sigma$ value, subject to constraints on both the mean and maximum temperature values to ensure feasible operating conditions. Additionally, constraints are imposed on the design variables to ensure the practicality and manufacturability of the optimized design. The problem formulation is presented in Tab.~\ref{tab:opt_form}.

\begin{table}[ht]
% \begin{spacing}{0.85}
\centering
\caption{\label{tab:opt_form} ESC optimization problem formulation}
\begin{tabular}{ccccc}

 & \textsc{Description}  & \textsc{Range (units)} & \textsc{Quantity} \\
 \hline
\textsc{Objective} & Temperature uniformity & $3\sigma_T$ $({{}^{\circ}\text{C}})$ & 1 \\ 
\hline
\textsc{Design variables} & Emboss contact ratio (inner zone)   & $ 0.01 \leq \text{CR1} \leq  0.1$  & 1\\ 

& Emboss contact ratio (outer zone)   & $ 0.01 \leq \text{CR2} \leq  0.1 $   & 1\\ 
& Inner coolant path height  & $ 5 \leq \text{H1} \leq 19.5 $  \text{(mm)} & 1\\ 
& Outer coolant path height   & $ 1 \leq \text{H2} \leq 19.5 $  \text{(mm)} & 1\\ 
& Inner coolant path width   & $ 5 \leq \text{W1} \leq 8 $  \text{(mm)} & 1\\ 
& Outer coolant path width & $ 5 \leq \text{W2} \leq 8 $  \text{(mm)} & 1\\ 
& Outer coolant path fin height   & $ 0 \leq \text{F1} \leq 10 $  \text{(mm)} & 1\\ 
 & & & Total design variables: $7$\\
 \hline 
\textsc{Constraints} 
 & Mean temperature & $\mu_T \leq 17$ $({{}^{\circ}\text{C}})$ & 1\\
 & Maximum temperature & $\text{max}(T) \leq 21.5$ $({{}^{\circ}\text{C}})$ & 1\\
& Design variables constraints& $\text{CR1} \leq \text{CR2} $ & 1\\
& & $\text{CR1} + \text{CR2} \leq 10 $ \text{(mm)} & 1\\
& & $\text{W2} \leq \text{W1} $ \text{(mm)} & 1\\

& & $\text{F1} \leq \text{W2} - 2 $ \text{(mm)} & 1\\
 & & & Total constraints: $6$\\

\end{tabular}

% \end{spacing}
\end{table}

\subsection{High- and low-fidelity models}

\begin{figure}[h]
    \centering
    \includegraphics[width = 15 cm]{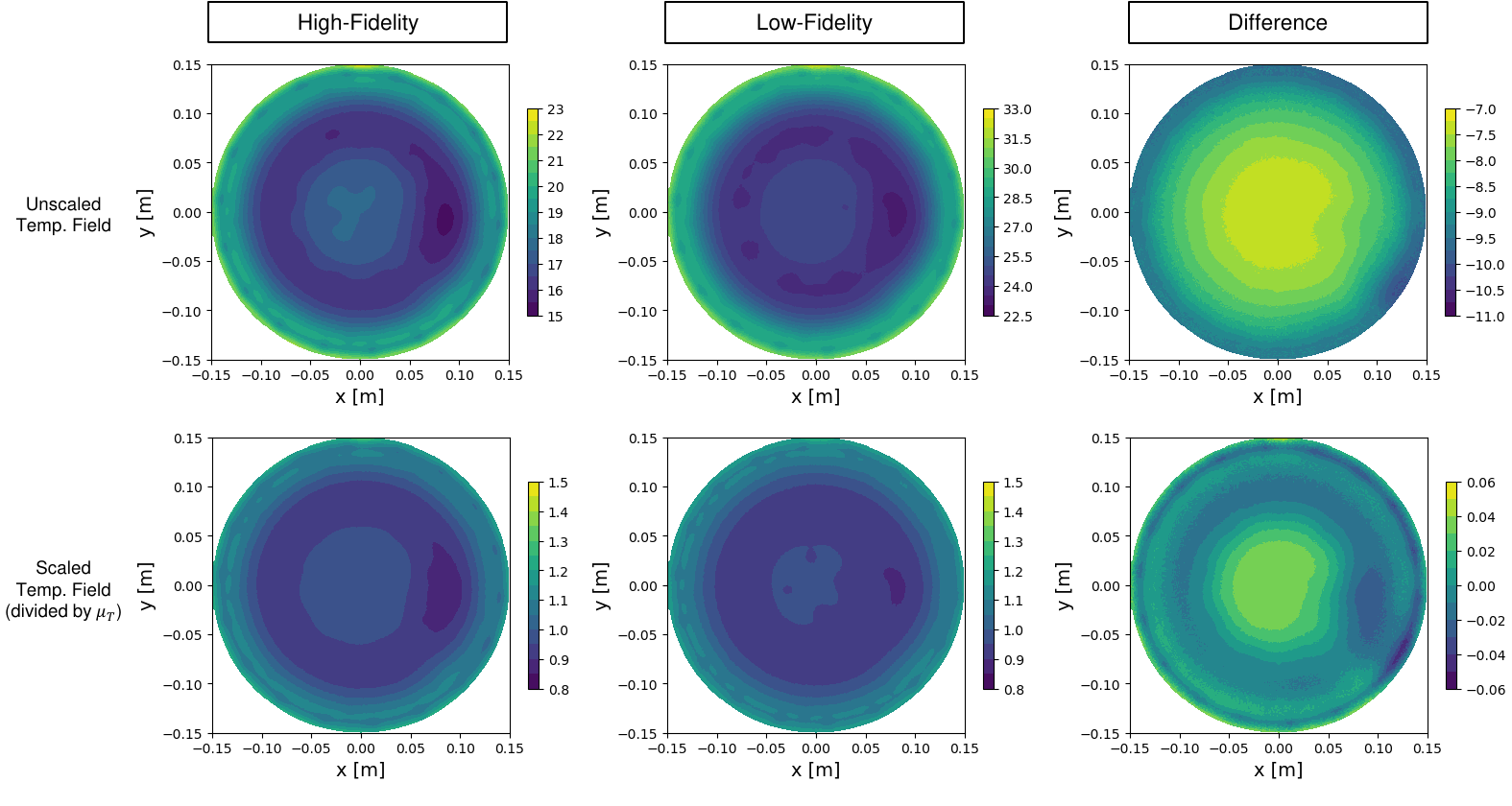}
    \caption{Comparison of LF and HF simulation results on the same design point. The LF simulation result demonstrates a good ability to capture the temperature uniformity of the field. 
}
    \label{fig:comp_LF_HF}
\end{figure}

For this problem, we leverage both high-fidelity (HF) and low-fidelity (LF) simulation models available through our industry partner. The HF model utilizes the Ansys Fluent Solver to predict the temperature field.
In this solver, a dynamic heat-transfer simulation model is coupled with a CFD model (SST k-omega turbulence model (\cite{menter1993zonal})) to compute the wafer temperature field.
While this model has been extensively validated for its accuracy, it comes with significant computational expense due to the complexity of the solver.
Conversely, the LF model uses the Ansys Mechanical Solver, which relies on a steady-state heat transfer model with simplified assumptions for the coolant flow. 
Although this approach is less precise in predicting the temperature field, it effectively captures the variation patterns of the temperature field with a much lower computational cost ($< 1/20$ of the HF simulation cost).
A detailed comparison between the HF and LF models is provided in Tab.~\ref{tab: Comp_LF_HF}, with a visual comparison of their simulation results shown in Fig.~\ref{fig:comp_LF_HF}. 
As seen in As shown in Fig.~\ref{fig:comp_LF_HF}, the LF simulation results differ significantly from the HF simulation across the entire temperature field, with an approximate 40\% discrepancy. However, the LF simulation still effectively captures the overall variation in the temperature field. This is evident when comparing the scaled temperature fields from both simulations: after normalizing by their respective mean temperatures, the maximum difference is shown to be just 6\%. This indicates that, despite the overall disparity, the LF simulation provides valuable insights into the temperature field's uniformity—one of the key quantities of interest in this optimization problem.

\begin{table}[]
\centering
\caption{Summary of high-fidelity and low-fidelity models.}
\begin{tabular}{c | c | c |c | c } 
% $9.8 \times 10^5$  
 Model type & Solver  &  No. of mesh nodes & No. of mesh nodes & Avg. eval. time\\ 
 & & for simulation & of the output (temp. field) & \\
 \hline
Low-fidelity  & Ansys Mechanical Solver & $\approx 9.9\times {10}^5$ & $\approx 2.5\times {10}^5$ & 44.27 s\\
High-fidelity & Ansys Fluent Solver & $\approx 7.1\times {10}^6$ & $\approx 9.0\times {10}^4$ & 919.4 s
\end{tabular}
\label{tab: Comp_LF_HF} 
\end{table}

% \begin{figure}%%
%     \centering
%     \subfloat[\centering HF simulation result ]{{\includegraphics[width=7 cm]{figures/HF_Sim.pdf} }}
%     \qquad
%     \subfloat[\centering LF simulation result]{{\includegraphics[width=7cm]{figures/LF_Sim.pdf}}}
%     \caption{Comparison of LF and HF simulation results on the same design point}%
%     \label{fig:comp_LF_HF}
% \end{figure}

% \begin{figure}%%
%     \centering
%     \subfloat[\centering HF simulation result]{{\includegraphics[width=7cm]{figures/LF_sim.jpg} }}
%     \qquad
%     \subfloat[\centering LF simulation result]{{\includegraphics[width=7cm]{figures/HF_sim.png}}
%     \caption{Comparison of LF and HF simulation results on the same design point }%
%     \label{fig:comp_LF_HF}
% \end{figure}

\subsection{Motivation and challenges for multi-fidelity surrogate modeling}
Combining simulation data from multiple levels of fidelity offers a compelling approach to solving complex optimization problems such as  the ESC design problem. By utilizing the complementary strengths of LF and HF simulations—where LF models capture broader trends at lower computational cost, and HF models provide precise but expensive simulations—we can enhance the accuracy of surrogate models while controlling computational expenses.
However, there are two primary challenges in constructing an effective multi-fidelity surrogate model for this problem:

\begin{enumerate}
    \item \textbf{Limited simulation data:} The high cost of running simulations means that we can only afford to generate a limited number of HF simulations (on the order of tens) and a moderate number of LF simulations (on the order of hundreds). This scarcity of data restricts the use of traditional machine learning-based surrogate models, which typically require significantly larger datasets to train effectively.

    \item \textbf{High-dimensional output space:} The surrogate model needs to predict the entire temperature field across more than 10,000 elements. This poses a substantial challenge, as classical surrogate modeling methods such as kriging and radial basis functions (RBF) struggle to handle such high-dimensional outputs directly.
\end{enumerate}
These challenges underscore the need for more sophisticated multi-fidelity surrogate modeling techniques capable of leveraging both HF and LF data effectively while addressing the high-dimensional nature of the problem. The combination of LF and HF data, despite its challenges, can lead to more accurate, computationally feasible models that capture both the large-scale variations and fine details of the temperature field.
% \subsection{Challenges}

% In this problem, there are three main challenges when it comes to building the surrogate models. Firstly, due to the expensive simulations, we can only afford a small amount of low-fidelity (100s) and high-fidelity (e.g. 10s) simulations. The small amount of data we have limits the use of machine learning-based surrogate modeling methods which reply a significantly larger amount of methods. Secondly, the surrogate model aims to predict the entire temperature field with more than 10,000 elements, in which traditionally surrogate modeling methods like kriging and radial basis functions cannot directly apply. Additionally, the mesh sizes related to low-fidelity and high-fidelity simulations are different, requiring the utilization of interpolation techniques when building multi-fidelity surrogate models.

\section{Methodology}
\label{Sec: Methodology}
% This paper proposes a 
This section describes a simple and efficient multi-fidelity surrogate modeling approach designed to accurately predict the temperature field on the wafer during the etching process, using a limited amount of low-fidelity and high-fidelity simulation data. The goal is to develop a surrogate model that can be used with numerical optimization methods to efficiently explore the design space and identify optimal parameters, while maintaining a balance between computational cost and predictive accuracy.

\subsection{Notations and problem set-up}

Let \( \mathbf{X} \subseteq \mathbb{R}^n \) represent the design parameter space, where \( n \) is the number of design variables. Let \( \mathbf{Y}_L \subseteq \mathbb{R}^{m_L} \) and \( \mathbf{Y}_H \subseteq \mathbb{R}^{m_H} \) denote the output spaces for the LF and HF simulation models, respectively.
We are provided with the following training data:
\begin{enumerate}
    \item \textbf{Low-fidelity training data}: \( \{ (\mathbf{x}_{L}^{(i)}, \mathbf{y}_{L}^{(i)}) \}_{i=1}^{N_L} \), where \( \mathbf{x}_{L}^{(i)} \in \mathbf{X} \) and \( \mathbf{y}_{L}^{(i)} \in \mathbf{Y}_L \), with \( N_L \) denoting the number of LF simulations.
    \item \textbf{High-fidelity training data}: \( \{ (\mathbf{x}_{H}^{(i)}, \mathbf{y}_{H}^{(i)}) \}_{i=1}^{N_H} \), where \( \mathbf{x}_{H}^{(i)} \in \mathbf{X} \) and \( \mathbf{y}_{H}^{(i)} \in \mathbf{Y}_H \), with \( N_H \) denoting the number of HF simulations.
\end{enumerate}
It is assumed that \( N_L > N_H \), and that the HF data points \( \{ \mathbf{x}_{H}^{(i)} \}_{i=1}^{N_H} \) are a subset of the LF data points \( \{ \mathbf{x}_{L}^{(i)} \}_{i=1}^{N_L} \), meaning that each HF data point has a corresponding LF data point at the same location in $\mathbf{X}$.

\subsection{Interpolation and dimension reduction}

In this problem, given that both LF and HF data consist of field data with more than 10,000 outputs, direct surrogate modeling with output dimension greater than 10,000 is extremely challenging when we are only given 100s of data. 
To address this, our strategy is to first apply dimension reduction techniques to project the high-dimensional outputs onto a much lower-dimensional latent space. 
% To address this, we first reduce the dimensionality of the data by projecting the high-dimensional outputs into a much lower-dimensional latent space. 
We use proper orthogonal decomposition (POD) for this task, as it efficiently captures the dominant modes of variation in the data, which is ideal for image-like datasets.

Our objective is to project both the LF and HF outputs onto the same latent space to facilitate multi-fidelity surrogate modeling. However, a key challenge is that the LF and HF data are originally defined on different  and non-standard grids. To address this, we perform an interpolation step to map both the LF and HF outputs onto a standard grid in Cartesian coordinates.
We can express the interpolation of the LF and HF data as follows:
\begin{equation}
\begin{aligned}
   &  \mathbf{y}_{L}^\prime = \mathcal{I}_{L \rightarrow {I}} (\mathbf{y}_L) \\
   & \mathbf{y}_H^\prime = \mathcal{I}_{H \rightarrow {I}} (\mathbf{y}_H)
\end{aligned}
\end{equation}
where  $\mathcal{I}_{L \rightarrow {I}}: \mathbb{R}^{m_L}  \to \mathbb{R}^{m_I} $ and $ \mathcal{I}_{H \rightarrow {I}}: \mathbb{R}^{m_H}  \to \mathbb{R}^{m_I} $ are the interpolation functions that transform the LF and HF outputs from their original grids to the standard grid, respectively. $m_{I}$ represents the total number of points in the standard grid.
The interpolation functions can be constructed based on various interpolation methods (e.g.,nearest neighbor, linear, inverse distance weighted etc.). 
% The general form is a linear combination of the LF data values, which adjusts the coarse-grid points to match the finer grid structure.
In this problem, we find that selecting a standard grid of 
$300 \times 300$ and applying the nearest neighbor interpolation method allows us to accurately interpolate both HF and LF temperature field outputs with minimal loss of information.

After the interpolations, we apply POD to project the interpolated outputs from LF and HF data onto the same latent variable space. This involves first assembling the LF and HF data in a matrix \( \mathbf{A} \in \mathbb{R}^{N \times m_I} \), where $ N = N_L + N_H$ is the number of snapshots (combined LF and HF data). We then apply singular value decomposition (SVD) to decompose the matrix as
\begin{equation}
    \mathbf{A} = \mathbf{U} \mathbf{\Sigma} \mathbf{V}^T
\end{equation}
where:
\( \mathbf{U} \in \mathbb{R}^{N \times N} \) is a unitary matrix containing the left singular vectors, which correspond to the temporal modes of the data,
\( \mathbf{\Sigma} \in \mathbb{R}^{N \times m_I} \) is a diagonal matrix of singular values, where the diagonal elements \( \sigma_1 \geq \sigma_2 \geq \dots \geq \sigma_N \geq 0 \) represent the significance (energy) of the corresponding modes,
\( \mathbf{V}^T \in \mathbb{R}^{m_I \times m_I} \) is a unitary matrix containing the right singular vectors, which correspond to the spatial modes (i.e., the dominant patterns in the field data).
The singular values in $ \mathbf{\Sigma} $ quantify the energy content of each mode. By retaining only the dominant modes (those with the largest singular values), we can reduce the dimensionality of the data while preserving most of the system’s important characteristics.
We select the top \( k \) singular values such that:
\begin{equation}
    \mathbf{A} \approx \mathbf{U}_k \mathbf{\Sigma}_k \mathbf{V}_k^T
\end{equation}
where \( \mathbf{U}_k \in \mathbb{R}^{N \times k} \) contains the first \( k \) left singular vectors, \( \mathbf{\Sigma}_k \in \mathbb{R}^{k \times k} \) contains the top \( k \) singular values, and \( \mathbf{V}_k^T \in \mathbb{R}^{k \times m_I} \) contains the first \( k \) right singular vectors.
By projecting the data onto the space spanned by the first \( k \) singular vectors, we obtain a lower-dimensional representation of the data. The reduced-dimension matrix \( \mathbf{V}_k \) captures the most important spatial features.

The steps of applying POD can be summarized as:
\begin{enumerate}
    \item Constructs a snapshot matrix by concatenating the state vectors corresponding to different points in the parameter space.
    \item Applies singular value decomposition (SVD) to the snapshot matrix.
    \item Identifies the low-dimensional subspace as the span of the leading $k$ right singular vectors.
\end{enumerate}

\subsection{Multi-fidelity surrogate modeling}
\begin{figure}[h]
    \centering
    \includegraphics[width = 14 cm]{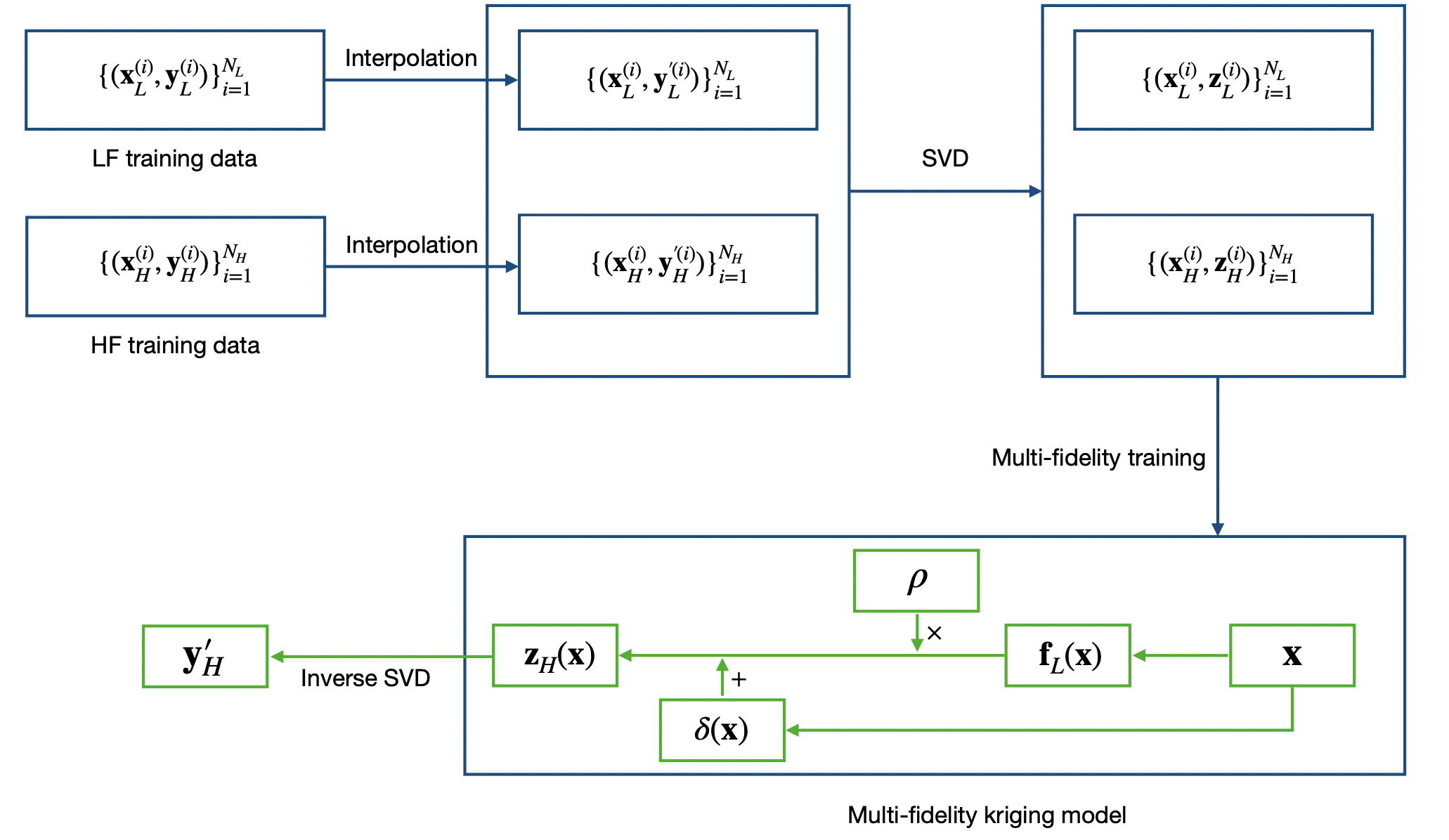}
    \caption{Flowchart for the proposed multi-fidelity surrogate modeling method.}
    \label{fig:flowchart_mf}
\end{figure}

Following the application of the POD, we project both the HF and low-fidelity LF output data onto a reduced 
k-dimensional latent space. This results in
\begin{equation}
\begin{aligned}
    &  \mathbf{y}_L^\prime \approx \mathbf{z}_L\mathbf{V}_k^T\\
    & \mathbf{y}_H^\prime \approx \mathbf{z}_H\mathbf{V}_k^T\\
\end{aligned}
\end{equation}
where $\mathbf{z}_L\in \mathbb{R}^{k} $ and $\mathbf{z}_H\in \mathbb{R}^{k} $ are the latent variables for the LF and HF output data, respectively.
With the reduced-dimensional data, we employ a multi-fidelity kriging method to combine the LF and HF datasets. Kriging is particularly well-suited for surrogate modeling in this scenario due to its effectiveness in handling small datasets (typically in the range of hundreds of data points) and low-dimensional output spaces (on the order of tens), which is the case after applying POD-based dimensionality reduction. Using the LF data pairs, \( \{(\mathbf{x}_{L}^{(i)}, \mathbf{z}_{L}^{(i)})\}_{i=1}^{N_L} \), and the HF data pairs, \( \{(\mathbf{x}_{H}^{(i)}, \mathbf{z}_{H}^{(i)})\}_{i=1}^{N_H} \), we train the multi-fidelity surrogate model as follows:
\begin{equation}
\mathbf{f}_{H}(\mathbf{x}): \Hat{\mathbf{z}}_{H}(\mathbf{x}) = \rho \, \mathbf{f}_{L}(\mathbf{x}) + \delta(\mathbf{x}) \quad \text{where}  \quad \mathbf{f}_{L}(\mathbf{x}):   \Hat{\mathbf{z}}_{L} = \mathbf{f}_L(\mathbf{x})
\end{equation}
where $\hat{\mathbf{z}}_{H} \in \mathbb{R}^{k} $ and $\hat{\mathbf{z}}_{L} \in \mathbb{R}^{k} $ represent the prediction vector of the HF and LF latent variables, respectively.
Here, \( \rho \) is a scaling constant, and \( \delta(\mathbf{x}) \) is the discrepancy function. The training of this multi-fidelity surrogate model proceeds in two main steps. First, the kriging surrogate model \( \hat{\mathbf{z}}_{L} = \mathbf{f}_L(\mathbf{x}) \) is trained using the LF data. Then, the constant \( \rho \) and the discrepancy function \( \delta(\mathbf{x}) \) (which is also modeled using kriging) are trained using the HF data and the predictions of the LF surrogate model at the HF data points. Detailed implementation of this method can be found in 
(\cite{le2013multi}).
This approach results in a multi-fidelity surrogate model that can predict the latent variable values for the HF simulation outputs, and we can easily recover the predicted temperature field data following
\begin{equation}
    \hat{\mathbf{y}}_H^\prime =\hat{\mathbf{z}}_H\mathbf{V}_k^T,
\end{equation}
where $\hat{\mathbf{y}}_H^\prime$ denotes the predicted temperature field on the standard grid.
The flowchart of this proposed method is shown in Fig.~\ref{fig:flowchart_mf}.

\section{Results}
\label{Sec: Results}
In this section, we compare the performance of three surrogate modeling methods:
\begin{enumerate}
    \item HF surrogate modeling: interpolation + POD + kriging model trained solely on HF data
    \item LF surrogate modeling: interpolation + POD + kriging model trained solely on LF data
    \item MF surrogate modeling (our proposed method): interpolation + POD + multi-fidelity kriging model trained on a combination of HF and LF data.
\end{enumerate}
To ensure comprehensive coverage of the design space, we generated 1500 design of experiments (DoE) points using the Latin Hypercube sampling method. These points span the entire input space, providing a robust dataset for both training and validation. We collected LF simulation results for all 1500 DoE points and HF simulation results for the first 150 DoE points. We trained all of the surrogate models through the Surrogate Modeling Toolbox (SMT) (\cite{SMT2019}) using the kriging with partial least squares (KPLS) method (\cite{bouhlel2016improving}).

\subsection{Surrogate modeling results}
To evaluate the accuracy of each surrogate modeling method, we randomly selected 30 out of the 150 available HF simulation results for validation. These 30 points were excluded from the training datasets and were used exclusively for testing the predictive performance of the surrogate models. Each surrogate modeling method was tested on three different datasets, each randomly selected from the remaining data, ensuring that no validation points were included in the training sets. This approach allowed us to rigorously assess the generalization capability of each model to new, unseen data points.
The accuracy of each surrogate model was measured by evaluating the predictions at the 30 validation points. For each surrogate modeling method, the \textit{root mean square error} (RMSE) was used to quantify the overall prediction error on the validation sets. In this case, the RMSE can be expressed as
\begin{equation}
    \text{RMSE} = \sqrt{\frac{1}{N_{\text{val}}} \sum_{i=1}^{N_{\text{val}}} \frac{1}{m_I} \sum_{j=1}^{m_I} \left(\mathbf{y}_H^{\prime(i,j)} - \hat{\mathbf{y}}_H^{\prime(i,j)} \right)^2}
\end{equation}
where $\mathbf{y}_H^{\prime(i,j)}$ represents the $j$-th element in the $i$-th temperature field output vector in the validation data set, and $N_{\text{val}}$ denotes the number of data in the validation set.
To obtain a robust measure of accuracy, for each surrogate modeling method. we calculated the \textit{average RMSE} across the surrogate models trained on the three different datasets.

\begin{figure}
    \centering
    \includegraphics[width= 0.5\linewidth]{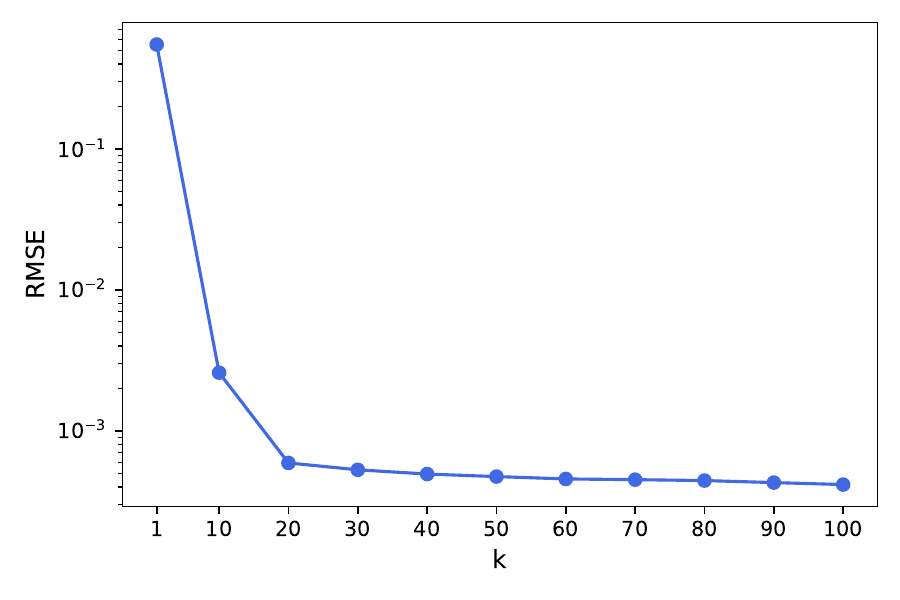}
    \caption{Reconstruction error vs latent variable size. Reconstruction error becomes insignificant after $k = 20$.}
    \label{fig:enter-label}
\end{figure}

\begin{figure}[h]
    \centering
    \includegraphics[width = 15 cm]{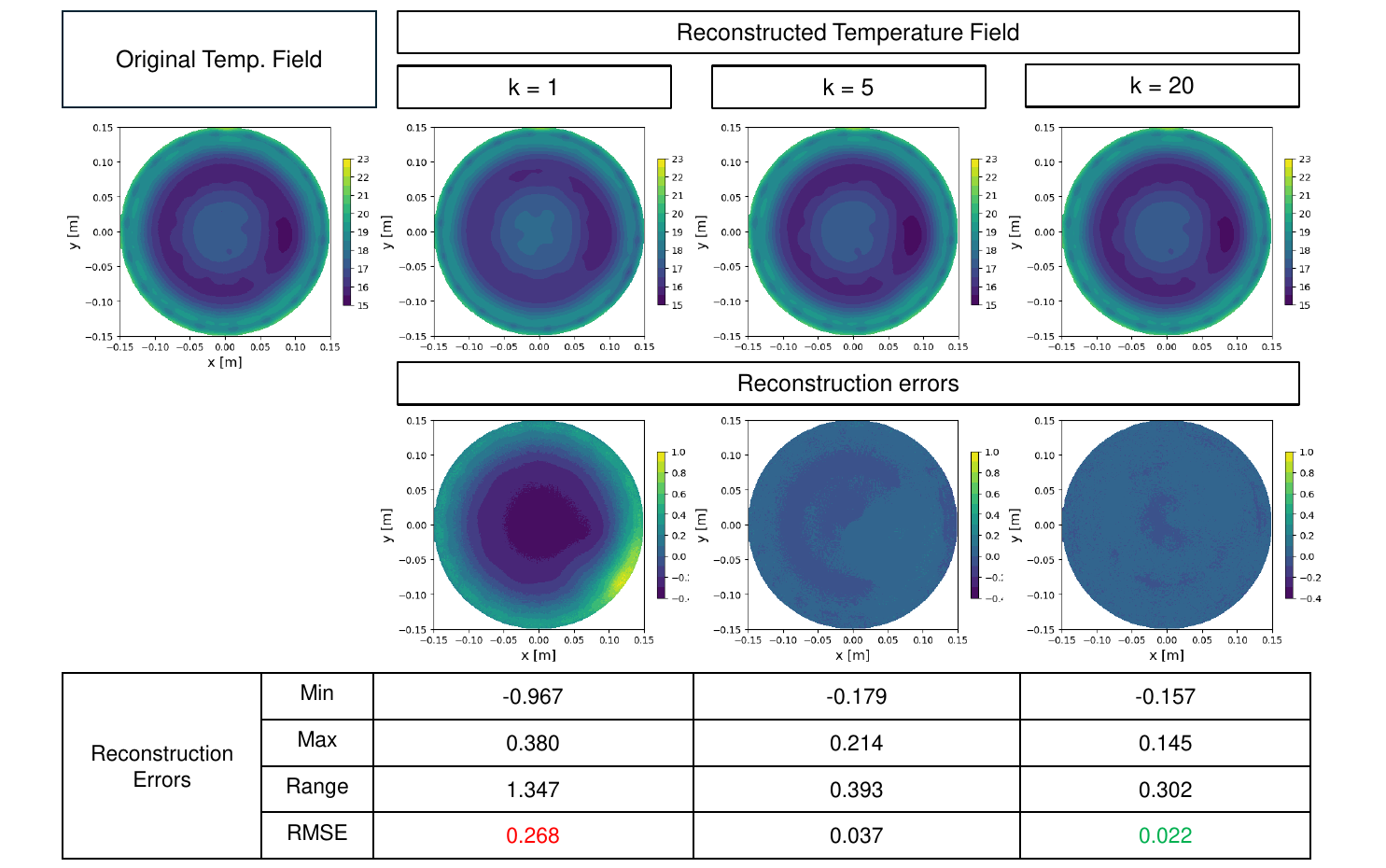}
    \caption{Temperature field predictions with various latent variable sizes. Significant improvements can be observed as we increase $k$ from 1 to 5 and from 5 to 20. 
}
    \label{fig:temp_field_recons}
\end{figure}

We first show the impact of the latent space dimension \( k \) on the accuracy of recovering the original temperature field through POD (Fig.~\ref{fig:enter-label}). 
Here, the RMSE is computed between the recovered temperature field and the original temperature field using the combined HF and LF training data.
As depicted in the figure, the RMSE decreases as the latent space dimension increases, almost reaching the lowest value at \( k = 20 \). Beyond this point, the RMSE remains relatively constant even with higher values of \( k \), indicating that increasing the dimensionality beyond \( k = 20 \) does not lead to much improvements in prediction accuracy.
Fig.~\ref{fig:temp_field_recons} shows a comparison between the original temperature field and the reconstructed temperature field using $k = 1$, 5, and 20, along with the reconstruction errors and their statistics. From this figure, we can observe significant improvements on the temperature field reconstruction errors as we increase the latent variable dimension from 1 to 5 and from 5 to 20, which confirms the effectiveness of the POD approach in reducing the dimension of the temperature field output.

% We first demonstrate the effect of the latent space dimension \( k \) on the accuracy of the surrogate models. The results for all three surrogate modeling methods are shown in Fig.~\ref{fig:rmse_vs_k}. As depicted in the figure, the average RMSE decreases as the latent variable size increases, almost reaching the lowest value at \( k = 10 \). Beyond this point, the RMSE remains relatively constant even with higher values of \( k \), indicating that increasing the dimensionality beyond \( k = 10 \) does not lead to much improvements in prediction accuracy.

\begin{figure}%%
    \centering
    \subfloat[\centering LF surrogate model ]{{\includegraphics[width=7cm]{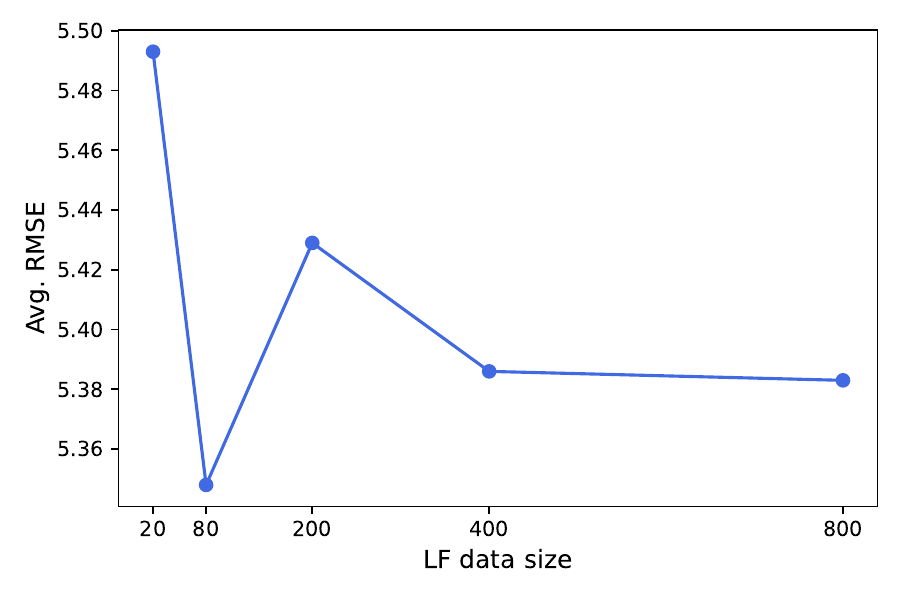}}}
    \quad 
    \subfloat[\centering HF surrogate model]{{\includegraphics[width=7cm]{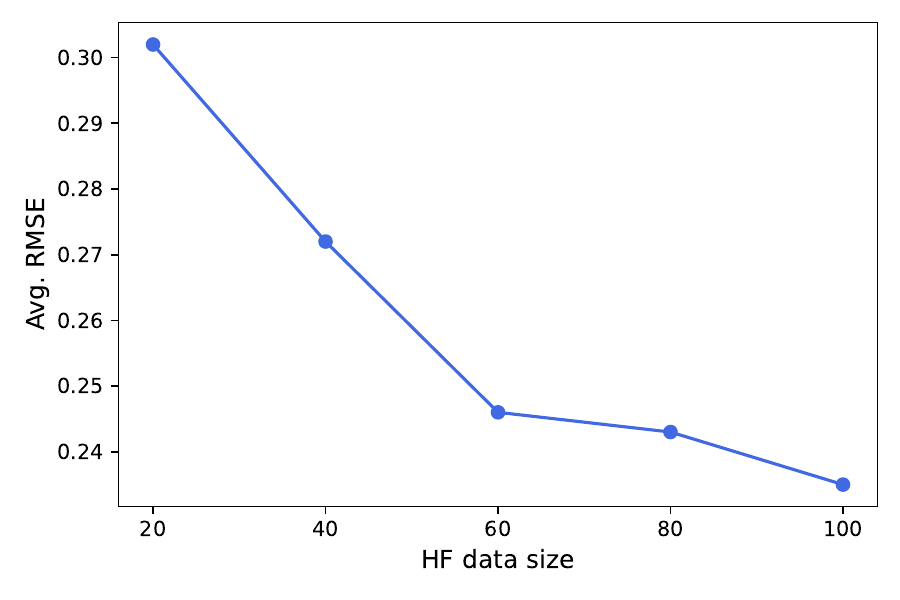}}}
    \caption{Average RMSE vs data size for LF and HF surrogate models ($k = 20$). The accuracy of the LF surrogate model does not increase with the data size while the HF surrogate model increases steadily.}%
    \label{fig:rmse_vs_data_size_LF_HF}
\end{figure}

\begin{figure}%%
    \centering
    \subfloat[\centering Average RMSE vs LF data size]{{\includegraphics[width=7 cm]{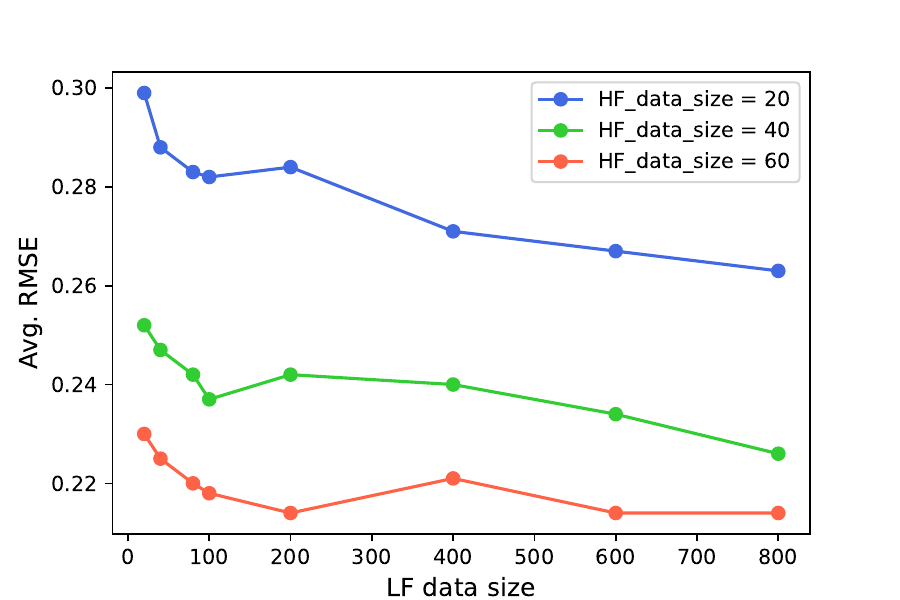} }}
    \qquad
    \subfloat[\centering Average RMSE vs HF data size]{{\includegraphics[width=7cm]{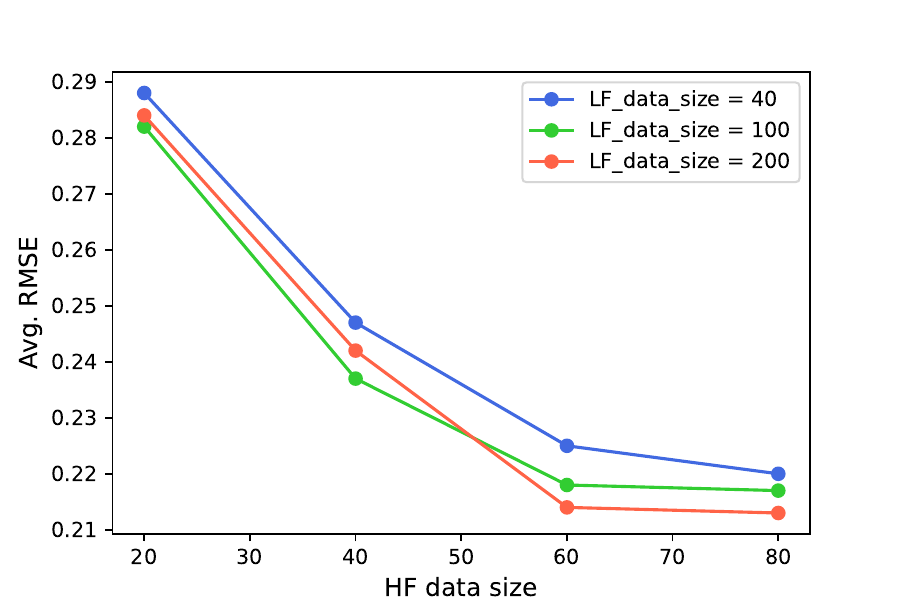}}}
    \caption{Average RMSE vs LF and HF data size for the MF surrogate model ($k = 20$). The accuracy of the MF surrogate model increases as we increase the LF and HF data sizes and is more sensitive towards the HF data size.}%
    \label{fig:rmse_vs_data_size_mf}
\end{figure}

\begin{figure}[h]
    \centering
    \includegraphics[width = 14 cm]{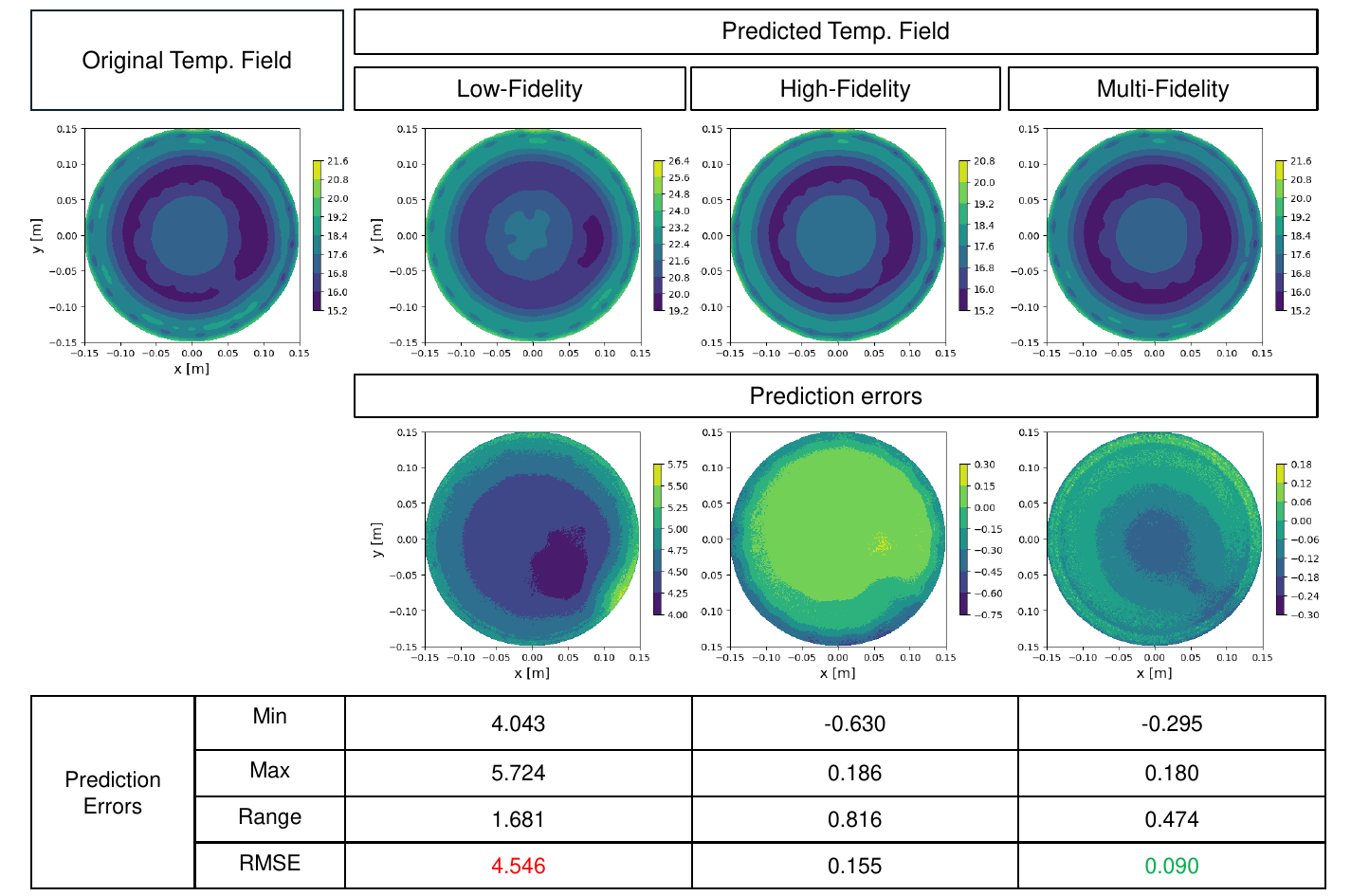}
    \caption{Temperature field predictions with different surrogate models (low-fidelity surrogate model with 400 data vs high-fidelity surrogate model with 80 data vs multi-fidelity surrogate model with 60 LF data + 100 HF data). MF surrogate model shows lower prediction errors the the HF and LF surrogate models.
}
    \label{fig:temp_field_predict}
\end{figure}

Following the previous observation, we selected \( k = 20 \) for further testing to investigate the effect of data size on the accuracy of each surrogate modeling method. The results for the LF and HF surrogate models are presented in Fig.~\ref{fig:rmse_vs_data_size_LF_HF}, while the results for the MF surrogate model are shown in Fig.~\ref{fig:rmse_vs_data_size_mf}.
In Fig.~\ref{fig:rmse_vs_data_size_LF_HF}, the results demonstrate that for the HF surrogate model, the RMSE consistently decreases as the data size increases. This trend indicates that larger HF datasets improve the model’s predictive accuracy. However, for the LF surrogate model, the increase in data size does not necessarily lead to better accuracy due to the inherent inaccuracy of the LF simulation data. This suggests that simply adding more LF data cannot overcome the limitations of the lower-fidelity model.
In contrast, Fig.~\ref{fig:rmse_vs_data_size_mf} shows that for the proposed MF surrogate model, increasing both LF and HF data sizes significantly improves the model’s accuracy. The RMSE decreases steadily as more LF and HF data are used, up to a point where the model reaches an accuracy limit. 
Additionally, the accuracy of the MF surrogate model is more sensitive towards the size of HF data than the LF data,  as HF data—serving as the ground truth—provides more valuable information. 
This highlights the effectiveness of combining LF and HF data in the MF surrogate model to achieve higher predictive performance compared to using LF or HF data alone.
Fig.~\ref{fig:temp_field_predict} presents the predicted temperature fields of the three surrogate models at a given design point, along with their respective prediction errors. The figure shows that the MF surrogate model, trained with 60 HF and 100 LF data points, achieves lower prediction errors than the HF surrogate model trained with 80 HF data points, while the LF surrogate model results in significantly higher prediction errors.

% \begin{figure}%%
%     \centering
%     \subfloat[\centering LF surrogate model with LF validation dataset]{{\includegraphics[width=7 cm]{figures/RMSE_vs_lf_data_size_lf_val.pdf} }}
%     \qquad
%     \subfloat[\centering LF surrogate model with HF validation dataset]{{\includegraphics[width=7cm]{figures/RMSE_vs_lf_data_size_hf_val.pdf}}}
%     \quad 
%     \subfloat[\centering HF surrogate model]{{\includegraphics[width=7cm]{figures/RMSE_vs_hf_data_size.pdf}}}
%     \caption{Average RMSE vs data size for LF and HF surrogate models ($k = 20$)}%
%     \label{fig:rmse_vs_data_size_LF_HF}
% \end{figure}

\begin{figure}%%
    \centering
    \subfloat[\centering Avg. RMSE, LF vs HF]{{\includegraphics[width=7 cm]{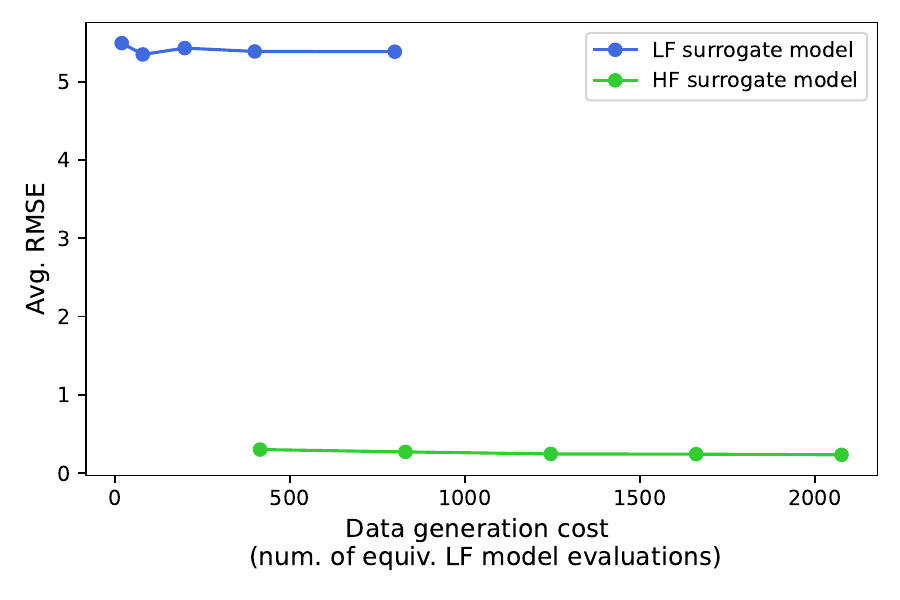} }}
    \qquad
    \subfloat[\centering   Avg. RMSE, HF vs MF]{{\includegraphics[width=7cm]{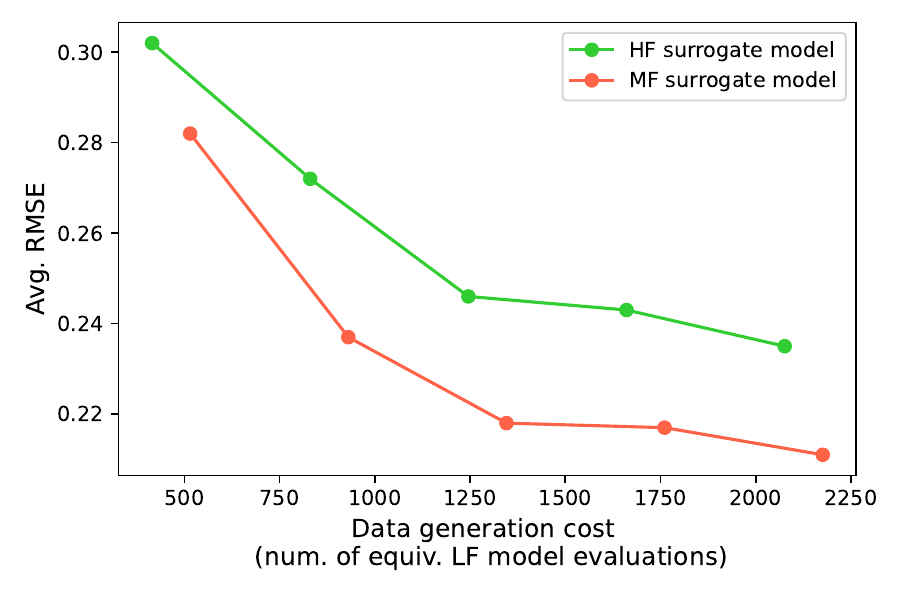}}} \\
    \subfloat[\centering Avg. relative errors on $max(T)$, LF vs HF]{{\includegraphics[width=7 cm]{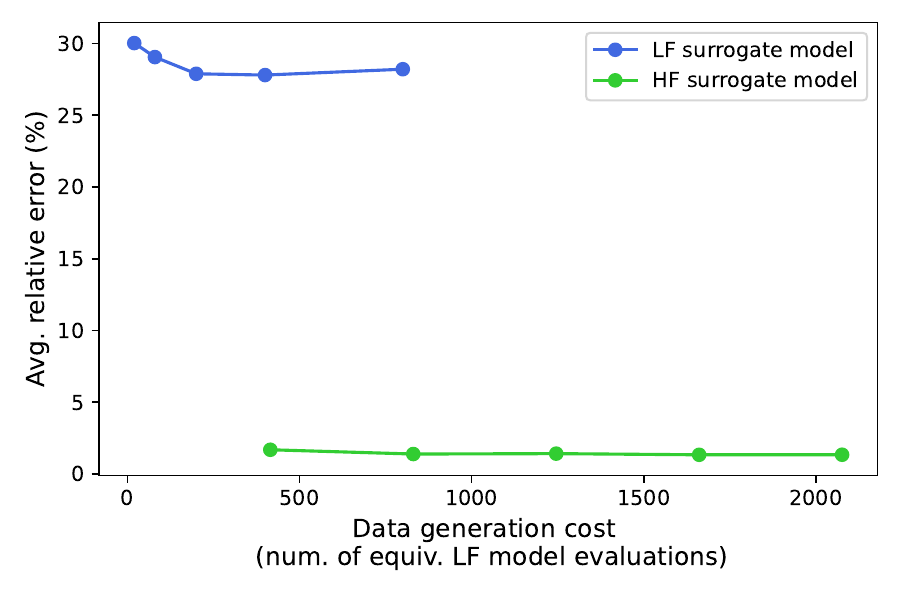} }}   
    \qquad
    \subfloat[\centering Avg. relative errors on $max(T)$, HF vs MF]{{\includegraphics[width=7 cm]{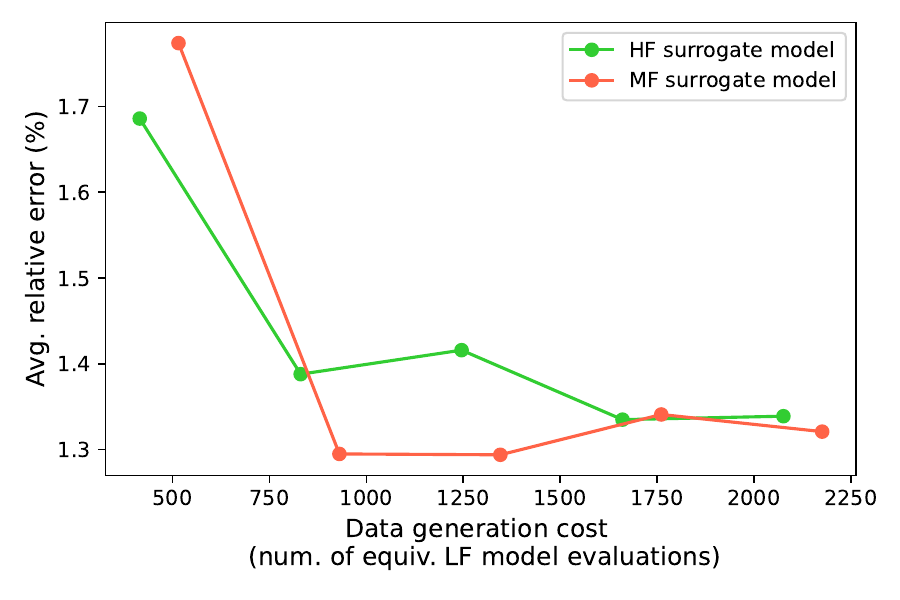} }} \\  
    \subfloat[\centering Avg. relative errors on $\mu_T$, LF vs HF]{{\includegraphics[width=7 cm]{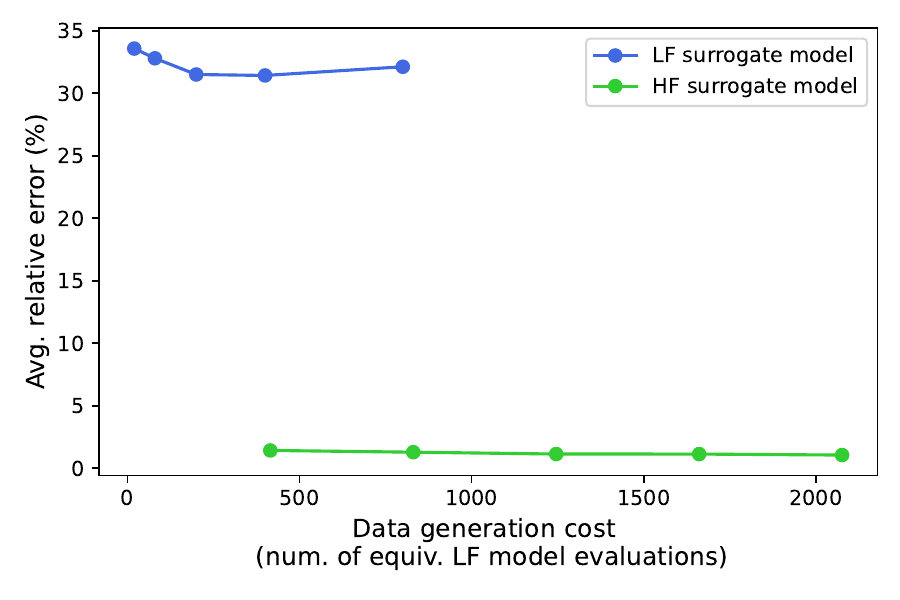} }}   
    \qquad
    \subfloat[\centering Avg. relative errors on $\mu_T$, HF vs MF]{{\includegraphics[width=7 cm]{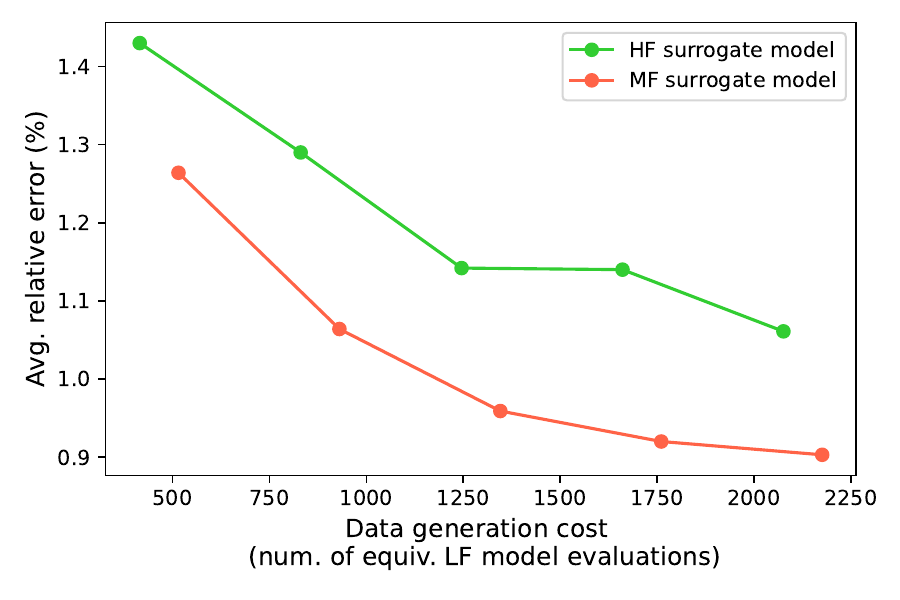} }} \\   
    \subfloat[\centering Avg. relative errors on $\sigma_T$, LF vs HF]{{\includegraphics[width=7 cm]{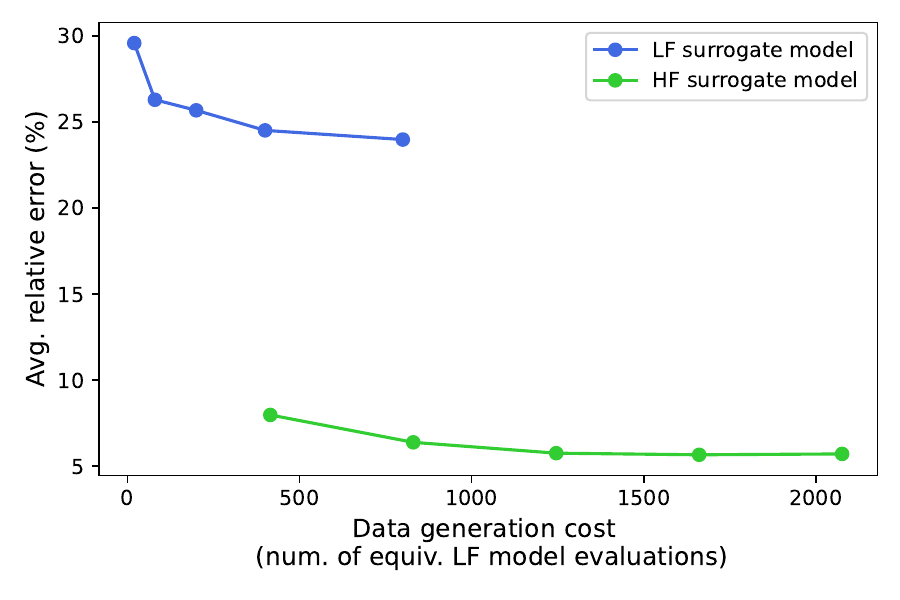} }}   
    \qquad
    \subfloat[\centering Avg. relative errors on $\sigma_T$, HF vs MF]{{\includegraphics[width=7 cm]{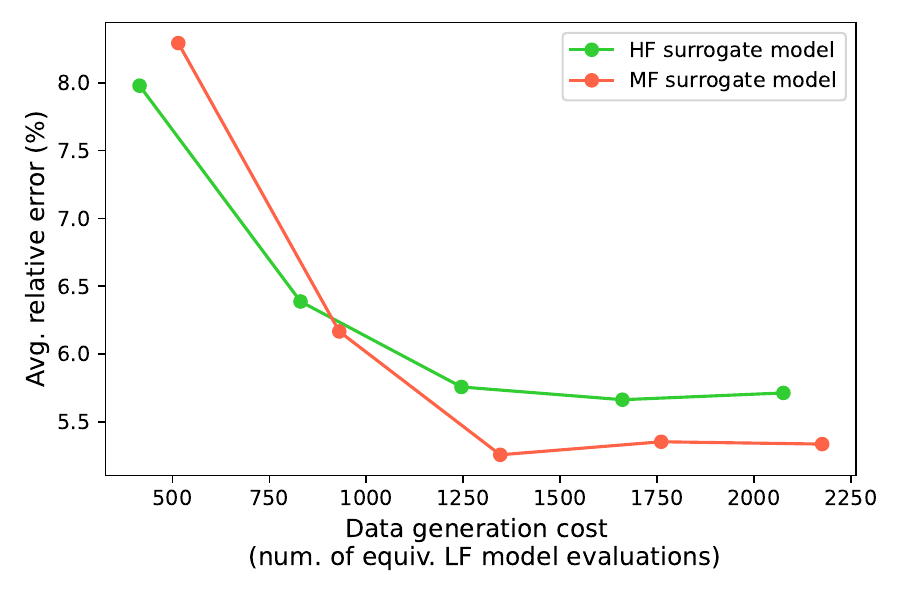} }} 
    \caption{Average RMSE and relative errors on QoIs vs data generation cost for LF, HF, and MF surrogate models (LF surrogate model trained with 20, 80, 200, 400, 800 LF data vs HF surrogate model trained with 20, 40, 60, 80, 100 HF data vs MF surrogate model trained with 20, 40, 60, 80, 100 HF data + 100 LF data). The MF surrogate model achieves higher accuracy than the LF and HF surrogate models under the same computational cost. }%
    \label{fig:convergence_plot}
\end{figure}

Lastly, we show convergence plots for the three surrogate modeling methods with respect to various computational costs (Fig.~\ref{fig:convergence_plot}). The computational cost is expressed in terms of the number of equivalent LF simulation evaluations to generate the training data, allowing for a fair comparison across methods with differing computational expenses. 
The results not only compare the average RMSE of each surrogate modeling method but also compare the average relative error in estimating the three QoIs in the ESC optimization problem: $\text{max}(T)$, $\mu_T$ and $\sigma_T$.
The results show that both the HF and MF surrogate models are significantly more accurate than the LF surrogate model across all metrics. 
Additionally, regarding the average RMSE, which is the most important metric to measure the accuracy of the surrogate model in predicting the entire temperature field, the MF surrogate model consistently outperforms the HF surrogate model, achieving approximately a 15\% reduction in error under the same computational cost.
Regarding the QoIs, the MF surrogate model performs similarly to the HF model in estimating the maximum temperature but surpasses the HF model in predicting the other two QoIs.

The results above show that, in this problem, by adding just 100 LF data points to the HF dataset,  our proposed method can significantly improve the accuracy of the HF kriging surrogate model. 
In the meantime, building the kriging surrogate model using solely LF data results in much lower accuracy,
The results demonstrate the efficiency of our proposed multi-fidelity approach in utilizing both LF and HF data to enhance predictive accuracy while maintaining lower computational demands compared to the HF surrogate model.

% \begin{figure}%
%     \centering
%     \subfloat[LF surrogate model with LF validation dataset]{{\includegraphics[width=7cm]{figures/rmse_vs_k_lf_lf_val.pdf}}}%
%     \qquad
%     \subfloat[LF surrogate model with HF validation dataset]{{\includegraphics[width=7cm]{figures/rmse_vs_k_lf_hf_val.pdf}}} \\
%     \subfloat[HF surrogate model]{{\includegraphics[width=7cm]{figures/rmse_vs_k_hf.pdf}}}%
%     \qquad
%     \subfloat[MF surrogate model]{{\includegraphics[width=7cm]{figures/rmse_vs_k_mf.pdf}}}%
%     \caption{Average RMSE vs latent variable size (LF surrogate model trained with 400 LF data vs HF surrogate model trained with 80 HF data vs MF surrogate model trained with 60 HF data  + 100 LF data)}%
%     \label{fig:rmse_vs_k}
% \end{figure}

\subsection{Optimization results}

\begin{table}[ht]
\centering
\caption{ Optimal designs (relative changes to the reference design) obtained from MF and HF surrogate models and their validation results. The best and second best solutions both come from using the MF surrogate models.}
\begin{tabular}{ccccccccc}

 & Ref. design   & \multicolumn{3}{c}{HF optimal designs} & \multicolumn{3}{c}{MF optimal designs}\\
  &    & Case 1 & Case 2 & Case 3 &Case 1 & Case 2 & Case 3\\
\hline 
\textbf{Design variables}: & \\
\text{CR1} & Redacted & +1.019  & +1.073 & -1.104& +1.023  & -0.143  & -0.996  \\
\text{CR2}  & Redacted & 
+1.881  & +1.893 & +4.004 & +1.876 & +2.043  & +3.876 \\
\text{H1}  & Redacted & -6.246 & -5.841  & -8.923 & -6.257 & -10.00 & -8.877   \\
\text{H2}  & Redacted & -3.169  & -3.752 & -5.538 & -3.164 & -3.389  & -5.689  \\
\text{W1} & Redacted & -2.698  & -2.625 & -2.685 & -2.702 & -2.644 & -2.678  \\
\text{W2}  & Redacted & -2.447  &  -2.367 & -2.619 & -2.451 & -2.696 & -2.573 \\
\text{F1}  & Redacted & +8.571 & +8.275 & +6.186 & +8.580 & +5.798 & +6.125\\
\hline
\textbf{Validation results}: & \\
$3 \sigma_T$ & 4.263  & 2.972  & 3.048 &	2.807 &	2.946 &	2.890 &	2.813\\
$\mu_T$ &  17.66  & 16.95  &	17.01 &	17.05 &	16.94  &	16.97 & 17.04
\\ 
$\text{max}(T)$ & 22.70 & 21.11 & 21.16  & 20.99 & 21.07 & 21.07 &	20.96
 \\
 \hline
% \textbf{Validation summary}: & \\
\begin{tabular}{c}
     \textbf{No. of violated opt. consts.}\\
     ($\mu_T > 17$ and/or $\text{max}(T) > 21.5$)
\end{tabular} & 2 & 0 & 1 & 1 & 0 & 0 & 1 \\
\hline
\textbf{Ranking of feasible designs} &   \\
(No constraints violation & - & 3rd & - & -  & 2nd & 1st & - \\
with minimum $3\sigma_T$) & \\
\end{tabular}
\label{tab:opt_results}
\end{table}

\begin{figure}
    \centering
    \includegraphics[width= 14 cm]{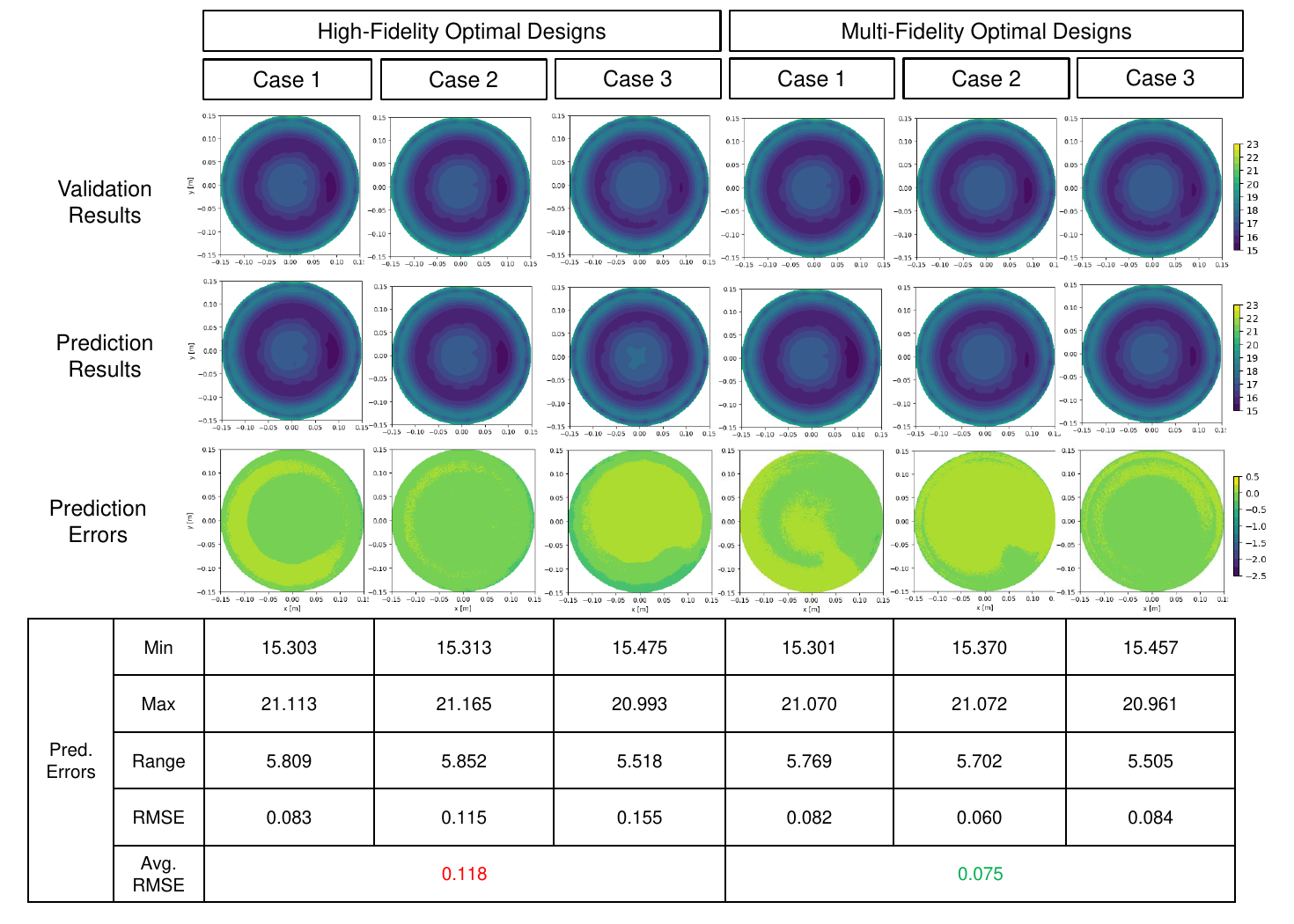}
    \caption{Prediction errors between the HF and MF optimal designs. MF surrogate models show significantly lower errors than the HF surrogate models. MF surrogate models show lower prediction errors than the HF surrogate models.}
    \label{fig:MF_HF_opt_pred}
\end{figure}

\begin{figure}
    \centering
    \includegraphics[width= 14 cm]{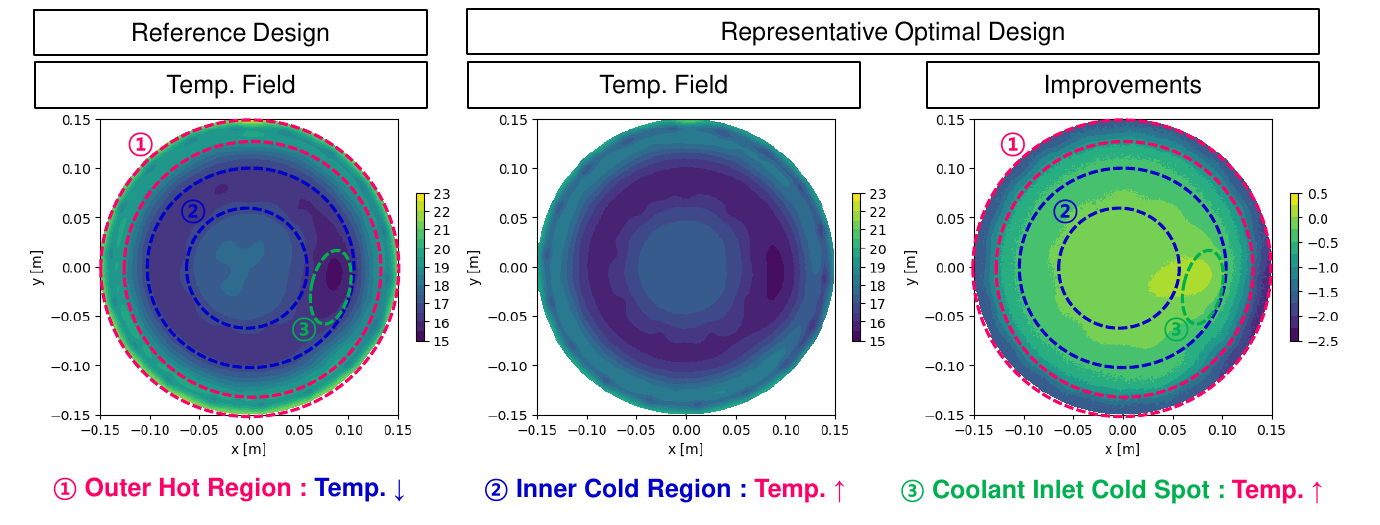}
    \caption{Demonstration of temperature field improvements on a representative optimal design. Significant improvement is observed on the uniformity of the temperature field.}
    \label{fig:rep_optimal_design}
\end{figure}

\begin{figure}
    \centering
    \includegraphics[width= 14 cm]{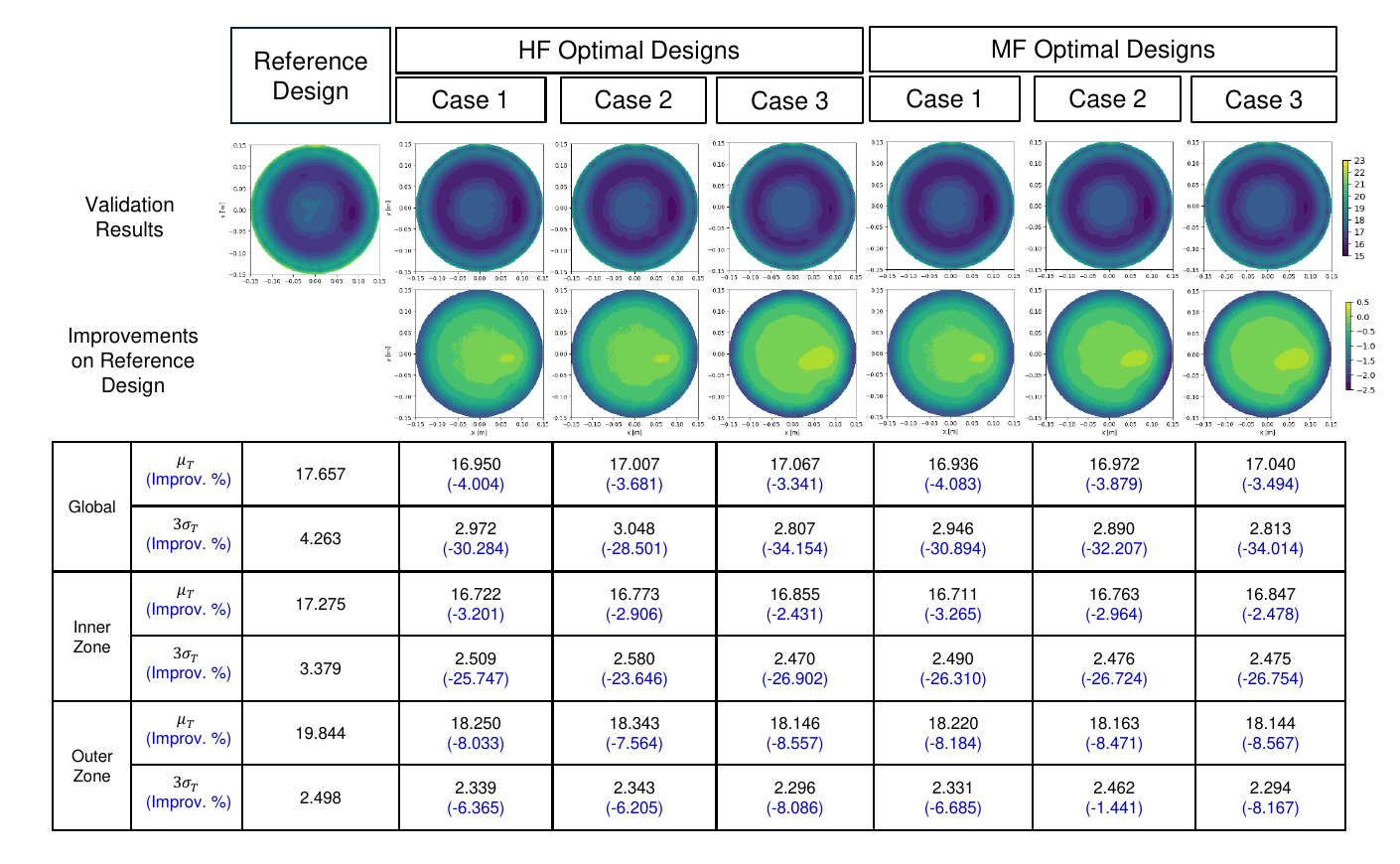}
    \caption{Comparison of temperature fields between the reference design and the optimal designs obtained through HF and MF surrogate models. Improvements in temperature uniformity are observed across all cases.}
    \label{fig:/HF_MF_opt_improvements}
\end{figure}

\begin{table}[]
\centering
\caption{Summary of optimization results. Overall, the MF surrogate models show better improvements on all of the QoIs while requiring lower data generation costs.}
\begin{tabular}{c  c  c  c c c c } 
 Surrogate model & Data size  & Avg. data generation cost &\multicolumn{3}{c}{Avg. improvement from} \\ 
& & (no. of LF evaluations) & \multicolumn{3}{c}{reference design}\\
& & & 3$\sigma_T$ & $\text{max}(T)$ & $\mu_T$ \\
 \hline
% Low-fidelity (LF) & 400 LF data& 400 & N/A \\
High-fidelity (HF) & 80 HF data & 1661 & 31.0\% & 7.11\% & 3.72\%\\
Multi-fidelity (MF) & 60 HF data + 100 LF data & 1346 & 32.4\% & 7.34\% &3.83\%\\
\end{tabular}
\label{tab: summary_opt} 
\end{table}

To evaluate the performance of each surrogate modeling method for the ESC device optimization problem, we test the effectiveness of these models in solving this optimization problem. 
We compare three different surrogate modeling approaches: HF surrogate modeling using 80 HF data points, LF surrogate modeling using 400 LF data points, and the MF surrogate modeling method using 60 HF and 100 LF data points.
Each surrogate model was trained on three randomly selected datasets, and the optimization problem, as defined in Tab.~\ref{tab:opt_results}, was solved using these models to obtain three optimal solutions for each method. All optimization problems were solved using the Sequential Least Squares Programming (SLSQP) optimizer from SciPy, with the same initial design for consistency.

To assess the performance of the optimal solutions, we sent these solutions to the industry partner for high-fidelity simulations (using the ground-truth model), which provided the key quantities of interest (QoIs) related to this optimization problem. 
For confidentiality reasons, the values of the design variables for the reference design have been redacted, and the optimal design values are presented as relative changes compared to the reference solutions.
The validation results are summarized in Tab.\ref{tab:opt_results}, and a comparison between the predicted and validation results for these optimal designs is presented in Fig.\ref{fig:MF_HF_opt_pred}.

As shown in Tab.~\ref{tab:opt_results}, in all three cases, both the HF and MF surrogate models produce optimal solutions that significantly improve upon the reference design for all three QoIs. However, validation results reveal that only three optimal designs satisfy all optimization constraints—one from the HF surrogate models and two from the MF surrogate models.
Among these three feasible solutions, the best and second-best (with the lowest objective function values) are generated using the MF surrogate models. This is attributed to the higher accuracy of the surrogate models built with the MF approach, as illustrated in Fig.~\ref{fig:MF_HF_opt_pred}, which shows lower prediction errors for the multi-fidelity optimal designs. 
% In general, more accurate surrogate models lead to better solutions in surrogate-based optimization problems, emphasizing the value of data-driven, accurate surrogate models in optimizing complex system designs.
This highlights the effectiveness of using accurate surrogate models built from data to optimize the design of complex systems.
% A comparison of temperature fields between optimal solutions and reference design on one case is shown in Fig.~\ref{fig:comp_temp_field}.

Regarding the improvements achieved by solving this optimization problem, Fig.~\ref{fig:rep_optimal_design} illustrates the temperature field change made by a representative optimal design. From this figure, we observe that by adjusting the dimensions of the coolant paths in both the inner and outer zones, along with other design variables, the optimal design results in a significant temperature decrease in the outer zone (the hotter region) of the wafer and a notable temperature increase in the inner zone (the cooler region).
Specifically, the largest temperature increase occurs at the coolant inlet's cold spot, which corresponds to the coldest area on the wafer. As a result, these adjustments have substantially improved the global uniformity of the temperature field, implying that solving this ESC optimization problem can effectively improve the performance of semiconductor manufacturing.

Fig.~\ref{fig:/HF_MF_opt_improvements} shows the temperature improvements over the reference designs for all cases using both the MF and HF surrogate models. The results indicate that while different datasets lead to distinct optimal solutions for both methods, improvements in temperature uniformity are clearly observed across all cases, with only minor differences between the MF and HF optimal solutions.
To assess the effectiveness of each method in solving the ESC design optimization problem, Table~\ref{tab: summary_opt} provides a summary of the overall optimization results, comparing data generation costs and average improvements achieved by the HF and MF surrogate models. The table shows that, on average, the MF surrogate models produced better optimal solutions with greater improvements across all three QoIs, while requiring approximately 20\% less data generation cost compared to the HF surrogate models.
% On average, the MF surrogate model, which was constructed using a lower data generation cost compared to the HF model, produced the better optimal solutions. 
This also proves the higher effectiveness of using our proposed surrogate modeling method rather than the standard HF kriging surrogate model in solving this optimization problem.
% It not only delivers superior performance in terms of the optimal solutions but also does so with reduced data generation costs, making it more efficient compared to the HF surrogate modeling method. Additionally, the failure of the LF surrogate model to converge in all cases underscores the importance of incorporating high-fidelity data to ensure the accuracy of the optimization process.

\section{Conclusion}
\label{Sec: Conclusion}

In this paper, we formulated a practical design optimization problem for a semiconductor manufacturing equipment, electrostatic chuck (ESC), addressing the challenge of improving temperature uniformity on the wafer surface. To tackle the difficulties posed by limited data and high-dimensional field outputs, we proposed a simple yet effective multi-fidelity surrogate modeling method that combines proper orthogonal decomposition (POD) and multi-fidelity kriging.

The proposed method leverages both low-fidelity (LF) and high-fidelity (HF) data to build a surrogate model that balances accuracy with computational cost. Our approach was tested to be 10\% more accurate than the kriging model using only HF data, under the same data generation cost. This demonstrates the strength of the multi-fidelity approach in extracting useful information from both LF and HF datasets while keeping the computational burden manageable.
Furthermore, the optimization results show that the MF surrogate model consistently produces better optimal solutions compared to the HF surrogate model, even though the MF models used requires approximately $20\%$ lower data generation costs. 
This highlights the potential of this multi-fidelity surrogate modeling  method as a cost-effective and accurate tool for solving the ESC design optimization problems in semiconductor manufacturing process.
This method can also be applied to other engineering design optimization problems, where only a limited amount of data is available to predict high-dimensional field outputs. 
Overall, the results demonstrate that the proposed multi-fidelity surrogate modeling method is not only computationally efficient but also highly effective in delivering great optimization performance for the ESC process problem, making it a valuable approach for similar engineering design problems.

For future work, a promising direction is to extend our proposed method to handle larger datasets. Currently, due to the limitations of the kriging method, the performance of our approach may degrade when dealing with thousands of input data points. 
Potentially, the machine learning-based multi-fidelity surrogate modeling methods may be better suited for solving this problem as we have more data to train the neural networks.
% A promising direction to address this challenge is to develop a more efficient multi-fidelity machine learning-based surrogate modeling approach capable of handling larger datasets. 
In the context of ESC design, we intend to build upon the results obtained in this study and furthur optimize the other key ESC design parameters, such as the shape of the coolant path, by leveraging physics-based simulation models. This optimization problem is more challenging (e.g. higher-dimensional design space) but can help further improve the performance of the semiconductor manufacturing process.

\section*{Declarations}

\subsection*{\textbf{Ethics approval and consent to participate}} 
Not applicable

\subsection*{\textbf{Consent for publication}}
Not applicable

\subsection*{\textbf{Funding}} 
This work was supported by Samsung Electronics Co., Ltd. (Fund number: IO231206-08178-01)

\subsection*{\textbf{Availability of data and materials}}
Not applicable

\subsection*{\textbf{Competing interests}}

The authors declare that they have no competing interests.

\subsection*{\textbf{Authors' contributions}}
B.W. and M.K. contributed equally to this work.  B.W., M.K. contributed on the conceptualization, methodology, software implementation, visualization, validation, and writing. T.Y., D.L., B.K., and D.S. contributed on data generation, and validation. J.H. contributed on conceptualization, methodology, writing, and funding acquisition. All authors reviewed the manuscript.

\subsection*{\textbf{Acknowledgements}}
Not applicable

\bibliographystyle{spbasic}      % basic style, author-year citations
\bibliography{bib}   % name your BibTeX data base

\begin{thebibliography}{28}
\providecommand{\natexlab}[1]{#1}
\providecommand{\url}[1]{{#1}}
\providecommand{\urlprefix}{URL }
\expandafter\ifx\csname urlstyle\endcsname\relax
  \providecommand{\doi}[1]{DOI~\discretionary{}{}{}#1}\else
  \providecommand{\doi}{DOI~\discretionary{}{}{}\begingroup \urlstyle{rm}\Url}\fi
\providecommand{\eprint}[2][]{\url{#2}}

\bibitem[{Achiam et~al.(2023)Achiam, Adler, Agarwal, Ahmad, Akkaya, Aleman, Almeida, Altenschmidt, Altman, Anadkat et~al.}]{achiam2023gpt}
Achiam J, Adler S, Agarwal S, Ahmad L, Akkaya I, Aleman FL, Almeida D, Altenschmidt J, Altman S, Anadkat S, et~al. (2023) Gpt-4 technical report. arXiv preprint arXiv:230308774

\bibitem[{Bouhlel et~al.(2016)Bouhlel, Bartoli, Otsmane, and Morlier}]{bouhlel2016improving}
Bouhlel MA, Bartoli N, Otsmane A, Morlier J (2016) Improving kriging surrogates of high-dimensional design models by partial least squares dimension reduction. Structural and Multidisciplinary Optimization 53:935--952

\bibitem[{Bouhlel et~al.(2019)Bouhlel, Hwang, Bartoli, Lafage, Morlier, and Martins}]{SMT2019}
Bouhlel MA, Hwang JT, Bartoli N, Lafage R, Morlier J, Martins JRRA (2019) A python surrogate modeling framework with derivatives. Advances in Engineering Software p 102662, \doi{https://doi.org/10.1016/j.advengsoft.2019.03.005}

\bibitem[{Bubenzer and Schmitt(1990)}]{bubenzer1990plasma}
Bubenzer A, Schmitt J (1990) Plasma processes under vacuum conditions. Vacuum 41(7-9):1957--1961

\bibitem[{Buhmann(2000)}]{buhmann2000radial}
Buhmann MD (2000) Radial basis functions. Acta numerica 9:1--38

\bibitem[{Chien et~al.(2011)Chien, Dauz{\`e}re-P{\'e}r{\`e}s, Ehm, Fowler, Jiang, Krishnaswamy, Lee, Moench, and Uzsoy}]{chien2011modelling}
Chien CF, Dauz{\`e}re-P{\'e}r{\`e}s S, Ehm H, Fowler JW, Jiang Z, Krishnaswamy S, Lee TE, Moench L, Uzsoy R (2011) Modelling and analysis of semiconductor manufacturing in a shrinking world: challenges and successes. European Journal of Industrial Engineering 4 5(3):254--271

\bibitem[{Fern{\'a}ndez-Godino(2016)}]{fernandez2016review}
Fern{\'a}ndez-Godino MG (2016) Review of multi-fidelity models. arXiv preprint arXiv:160907196

\bibitem[{Hwang and Martins(2018)}]{hwang2018fast}
Hwang JT, Martins JR (2018) A fast-prediction surrogate model for large datasets. Aerospace Science and Technology 75:74--87

\bibitem[{Jin et~al.(2024)Jin, Lim, Zhao, Mamunuru, Roham, and Gu}]{jin2024machine}
Jin Z, Lim DD, Zhao X, Mamunuru M, Roham S, Gu GX (2024) Machine learning enabled optimization of showerhead design for semiconductor deposition process. Journal of Intelligent Manufacturing 35(2):925--935

\bibitem[{Kennedy and O'Hagan(2001)}]{kennedy2001bayesian}
Kennedy MC, O'Hagan A (2001) Bayesian calibration of computer models. Journal of the Royal Statistical Society: Series B (Statistical Methodology) 63(3):425--464

\bibitem[{Klick and Bernt(2006)}]{klick2006microscopic}
Klick M, Bernt M (2006) Microscopic approach to an equation for the heat flow between wafer and e-chuck. Journal of Vacuum Science \& Technology B: Microelectronics and Nanometer Structures Processing, Measurement, and Phenomena 24(6):2509--2517

\bibitem[{Krige(1951)}]{krige1951statistical}
Krige DG (1951) A statistical approach to some basic mine valuation problems on the witwatersrand. Journal of the Southern African Institute of Mining and Metallurgy 52(6):119--139

\bibitem[{Le~Gratiet(2013)}]{le2013multi}
Le~Gratiet L (2013) Multi-fidelity gaussian process regression for computer experiments. PhD thesis, Universit{\'e} Paris-Diderot-Paris VII

\bibitem[{Li et~al.(2022)Li, Kou, and Zhang}]{li2022deep}
Li K, Kou J, Zhang W (2022) Deep learning for multifidelity aerodynamic distribution modeling from experimental and simulation data. AIAA Journal 60(7):4413--4427

\bibitem[{Liao et~al.(2018)Liao, Hsiau, and Chuang}]{liao2018modeling}
Liao C, Hsiau S, Chuang T (2018) Modeling and designing a new gas injection diffusion system for metalorganic chemical vapor deposition. Heat and Mass Transfer 54:115--123

\bibitem[{Matheron(1963)}]{matheron1963principles}
Matheron G (1963) Principles of geostatistics. Economic geology 58(8):1246--1266

\bibitem[{May and Spanos(2006)}]{may2006fundamentals}
May GS, Spanos CJ (2006) Fundamentals of semiconductor manufacturing and process control. John Wiley \& Sons

\bibitem[{Menter(1993)}]{menter1993zonal}
Menter F (1993) Zonal two equation kw turbulence models for aerodynamic flows. In: 23rd fluid dynamics, plasmadynamics, and lasers conference, p 2906

\bibitem[{O’Hanlon and Parks(1992)}]{o1992impact}
O’Hanlon JF, Parks HG (1992) Impact of vacuum equipment contamination on semiconductor yield. Journal of Vacuum Science \& Technology A: Vacuum, Surfaces, and Films 10(4):1863--1868

\bibitem[{Peherstorfer et~al.(2018)Peherstorfer, Willcox, and Gunzburger}]{peherstorfer2018survey}
Peherstorfer B, Willcox K, Gunzburger M (2018) Survey of multifidelity methods in uncertainty propagation, inference, and optimization. Siam Review 60(3):550--591

\bibitem[{Rumelhart et~al.(1986)Rumelhart, Hinton, and Williams}]{rumelhart1986learning}
Rumelhart DE, Hinton GE, Williams RJ (1986) Learning representations by back-propagating errors. nature 323(6088):533--536

\bibitem[{Shen et~al.(2024)Shen, Patel, Xu, and Alonso}]{shen2024application}
Shen Y, Patel HC, Xu Z, Alonso JJ (2024) Application of multi-fidelity transfer learning with autoencoders for efficient construction of surrogate models. In: AIAA SCITECH 2024 Forum, p 0013

\bibitem[{Shepard(1968)}]{shepard1968two}
Shepard D (1968) A two-dimensional interpolation function for irregularly-spaced data. In: Proceedings of the 1968 23rd ACM national conference, pp 517--524

\bibitem[{Sun et~al.(2020)Sun, Gao, Pan, and Wang}]{sun2020surrogate}
Sun L, Gao H, Pan S, Wang JX (2020) Surrogate modeling for fluid flows based on physics-constrained deep learning without simulation data. Computer Methods in Applied Mechanics and Engineering 361:112732

\bibitem[{Voulodimos et~al.(2018)Voulodimos, Doulamis, Doulamis, and Protopapadakis}]{voulodimos2018deep}
Voulodimos A, Doulamis N, Doulamis A, Protopapadakis E (2018) Deep learning for computer vision: A brief review. Computational intelligence and neuroscience 2018(1):7068349

\bibitem[{Wong et~al.(2020)Wong, Akarvardar, Antoniadis, Bokor, Hu, King-Liu, Mitra, Plummer, and Salahuddin}]{wong2020density}
Wong HSP, Akarvardar K, Antoniadis D, Bokor J, Hu C, King-Liu TJ, Mitra S, Plummer JD, Salahuddin S (2020) A density metric for semiconductor technology [point of view]. Proceedings of the IEEE 108(4):478--482

\bibitem[{Yoon et~al.(2023)Yoon, Cho, Choi, and Hong}]{yoon2023heat}
Yoon TW, Cho SI, Choi M, Hong SJ (2023) Heat transfer mechanism of electrostatic chuck surface and wafer backside to improve wafer temperature uniformity. Journal of Vacuum Science \& Technology B 41(4)

\bibitem[{Youn and Hong(2024)}]{youn2024enhanced}
Youn JH, Hong SJ (2024) Enhanced temperature uniformity of electrostatic chuck: ceramic surface contact ratio and backside gas pressure. Japanese Journal of Applied Physics 63(4):04SP72

\end{thebibliography}

\end{document}